# Design and demonstration of a direct air capture system with moisture-driven CO$_2$ delivery into aqueous medium

Justin Flory[a], Samantha Taylor[a], Shuqin Li[b], Sunil Tiwari[a], Garrett Cole[c], Amory Lowe[a], Lindsey Hamblin[a], Samuel Piorkowski[b], Matthew Ryan[a], Thiago Stangherlin Barbosa[a], Jason Kmon[a], Nick Lowery[a], Joel Eliston[a], Jason C. Quinn[c], John McGowen[d], Matthew D. Green[a], Klaus Lackner[a], Wim Vermaas[b]

[a]Center for Negative Carbon Emissions at Arizona State University (ASU), 777 E. University Dr., Tempe, Arizona, 85281, USA
[b]School of Life Sciences at ASU, Tempe, Arizona, 85287-4501, USA
[c]Sustainability Sciences LLC, Steamboat Springs, Colorado 80487, USA
[d]Center for Algae Technology and Innovation (AzCATI) at ASU, 7418 E. Innovation Way South, Mesa, Arizona, 85234, USA

**Abstract**

A moisture-driven air capture (DAC) system was designed and demonstrated, consisting of anion-exchange resin (AER) sorbent beads or sheets circulated between dry ambient air, where the sorbent dried and CO$_2$ was captured, and an alkaline aqueous medium, where captured CO$_2$ was delivered. A laboratory-scale system delivering ~1 g CO$_2$ per day was demonstrated in a laminar flow hood and a small pilot-scale system that could deliver ~100 g CO$_2$ daily was operated outdoors in a 4.2 m$^2$ (areal surface area) raceway pond. Elongated mesh tube packets were designed to contain AER beads with high surface area for contacting the air and were found to reduce drying and CO$_2$ loading time ~4-fold over larger mesh bags. A model was developed based on the sorbent drying and CO$_2$ loading rate data to accurately estimate the quantity of CO$_2$ delivered by the system based on wind speeds recorded by a nearby weather station. Whereas this system was designed for CO$_2$ delivery for cultivating photosynthetic microbes, its potential uses are much broader and include CO$_2$ use in the food and beverage industry, conversion to fuels and chemicals, and sequestration. Techno-economic assessments for a practical scenario based on current results are $670/tonne to capture CO$_2$ into an alkaline solution and an additional $280/tonne to extract CO$_2$ from solution, purify and compress to 15 MPa for sequestration. An aspirational scenario modelling reasonable improvements to develop AER sorbents with a capacity of 4 mmol CO$_2$ per gram of sorbent and water uptake of 50 wt.%, which leads to sorbent drying and loading within 1 h, shows a potential to reach $51/tonne to capture CO$_2$ into an alkaline solution and an additional $109/tonne to get to 15 MPa for sequestration. Life cycle analysis shows the aspirational moisture-driven process uses up to 87% less energy than thermal and/or vacuum swing DAC by using energy from water evaporation; however, ~330 wt.% water uptake by the sorbent contained in a hydrophilic mesh packets leads to ~33-fold higher water use than the thermodynamic limits, which emphasizes future research is needed to increase sorbent hydrophobicity while maintaining and further increasing ion exchange capacity needed to bind CO$_2$.

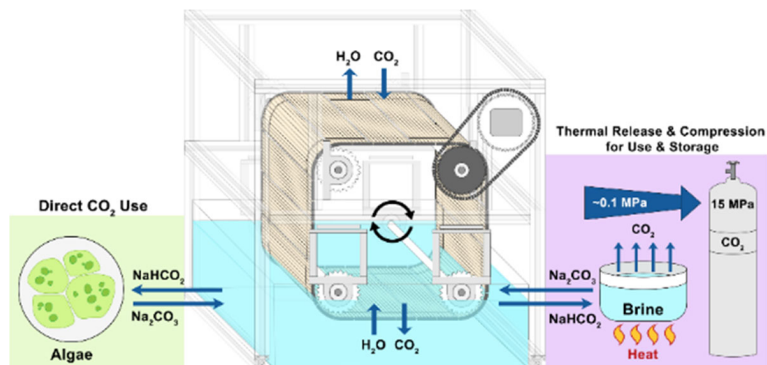

**Graphical Abstract.** Schematic of moisture-driven direct air capture system with CO$_2$ delivery into aqueous medium for direct use to cultivate algae or thermal extraction and compression for use, conversion or storage.

**Terms and abbreviations**

| | |
|---|---|
| AER | Anion Exchange Resin |
| AzCATI | Arizona Center for Algae Technology and Innovation |
| DI | Deionized |
| DAC | Direct Air Capture |
| IC | Inorganic Carbon |
| RH | Relative Humidity |

**Highlights**

- Outdoor demonstration of a moisture-driven DAC system with techno-economic analysis showing a pathway to $51/tonne into alkaline solutions for cultivating microalgae and mineral sequestration and $160/tonne $CO_2$ at 15 MPa suitable for use, conversion or geologic storage
- Elongated mesh tube packet design for containing beaded sorbents maintains fast $CO_2$ release and significantly decreases drying and loading times by about four-fold over mesh bags

**Introduction**

Balancing the world's carbon budget to stop the rise of the atmospheric carbon dioxide ($CO_2$) level, which is the dominant cause of global warming, requires both immense $CO_2$ emission reductions and gigaton-scale carbon dioxide removal (CDR) to address hard-to-abate industrial sectors and potentially remove past emissions.[1,2] Current CDR technologies, such as direct air capture (DAC), require substantial innovations to bring energy use and costs down by an order of magnitude.[3] Most DAC technologies use liquid or solid sorbent materials with strong binding affinities to capture $CO_2$ directly from atmospheric air to produce concentrated $CO_2$ streams suitable for sequestration in geologic formations, minerals and building materials, conversion into sustainable fuels or chemicals, or enhancing growth of plants (in greenhouses), microalgae or cyanobacteria.[4] While the thermodynamic energy requirement for $CO_2$ desorption from conventional sorbents may be around 50 kJ/mol $CO_2$ (315 kWh/tonne $CO_2$) on average,[5–7] real-world temperature-vacuum swing DAC systems typically require 1,800–4,000 kWh/tonne $CO_2$[8–10] due to sorbent heat capacity, system inefficiencies and low $CO_2$ partial pressures. Thus, the inherent energy costs of moving large volumes of air and heat and/or vacuum pressure to overcome the strong binding affinity of $CO_2$ to the sorbent material[3] hinder their scale-up and deployment.[11]

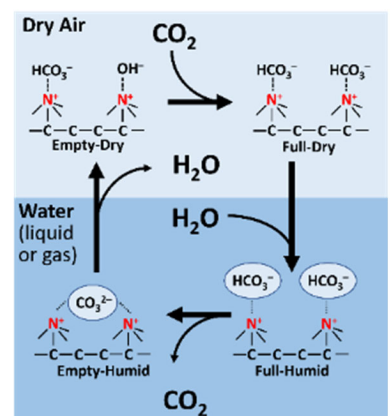

**Figure 1.** Moisture-driven direct air capture mechanism in strong-base anion-exchange resins (AER).

Passive $CO_2$ capture from ambient airflows eliminates costs and energy use from fans by removing only small fractions of $CO_2$ to maintain near ambient $CO_2$ concentrations at the sorbent interface[5] or using enhanced flows around physical structures.[12,13] However, passive capture systems require very low-pressure-drop contactors[13] to achieve practical $CO_2$ loading rates from ambient airflows and greater physically distributed $CO_2$ capture infrastructure.[5,14,15] Moisture-driven DAC leverages changes in hydration surrounding the $CO_2$ capture sites within anion-exchange resins (AER) to drive $CO_2$ release without external energy inputs, due to properties of water confined within nano-porous structures within the sorbent.[16–19] Wang, Lackner and Wright first showed that strong-base AERs containing fixed cations (e.g., quaternary ammonium; $NR_4^+$) that are charge balanced by bicarbonate ($HCO_3^-$), carbonate ($CO_3^{2-}$) or hydroxide ($OH^-$) anions capture $CO_2$ from ambient air when dry and release the $CO_2$ when wet to increase the partial pressure of $CO_2$ up to 500-fold through a moisture-driven process as shown in **Figure 1**.[20–22] When the AER dries, $CO_3^{2-}$ abstracts a proton ($H^+$) from a nearby water molecule to form $HCO_3^-$ and the residual $OH^-$ associates with the other $NR_4^+$ to maintain charge neutrality. The OH⁻ then captures $CO_2$ from ambient air to form $HCO_3^-$. When the AER is then exposed to liquid water or vapor, $CO_2$ is released just as it would from an aqueous $HCO_3^-$ solution, and the $CO_3^{2-}$ remaining on the AER restarts the capture cycle after drying.

Significant progress has been made to advance moisture swing DAC,[23–28] but little work has been done to evaluate these systems under expected operating conditions, leaving significant technical risks associated with their scale-up.[29] Here, we describe a moisture-driven DAC and $CO_2$ delivery system that has been designed and demonstrated to capture $CO_2$ passively from natural wind flows from any direction using commercially available anion exchange resin beads. The process also delivers $CO_2$ into aqueous alkaline medium, which retains the captured $CO_2$ without loss to the environment, and releases $CO_2$ from the sorbent without the need for external heat or vacuum energy. While the crude $CO_2$ product contained in the alkaline medium is suitable for cultivating photosynthetic microbes or mineralization, a process is modeled for thermally extracting $CO_2$ from an aqueous alkaline brine and compressing the $CO_2$ to 15 MPa, suitable for geologic sequestration, to compare the costs and energy use with other DAC systems.

## 1. Materials and Methods

Theory of operation: Based on the principles of moisture-driven DAC,[5,20,22] we designed a conveyance system (**Figure 2**) that rotates a moisture-driven sorbent belt attached to parallel drive chains between a $CO_2$ capture zone—where the sorbent is exposed to dry ambient air to capture $CO_2$ using natural wind flows—and a $CO_2$ release zone where the sorbent is immersed an aqueous medium and $CO_2$ is released.[30] The aqueous medium contains a sufficient level of alkalinity to retain and store the $CO_2$ as inorganic carbon (IC) in the form of $HCO_3^-$ to prevent release back into the environment. The IC concentration is suitable for accelerating the growth of photosynthetic microbes (microalgae and cyanobacteria), forming stable carbonate minerals, or extracting into a concentrated $CO_2$ gas stream using heat or acidification to further concentrate and compress the $CO_2$ for other utilization or sequestration applications.

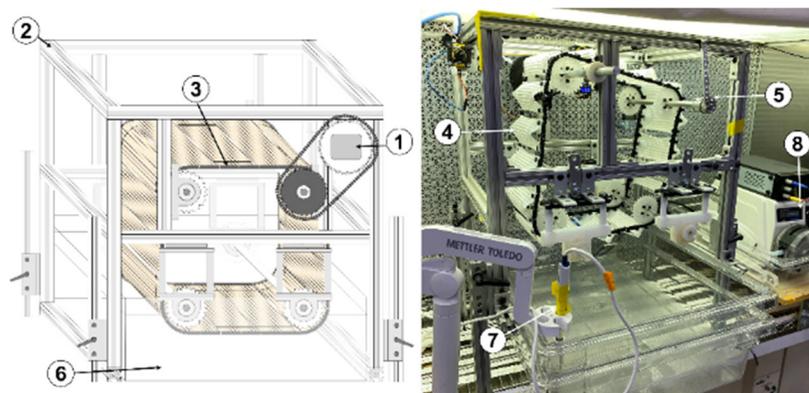

**Figure 2.** Schematic (left) and photo (right) of the *1g* system installed in a laminar flow hood. All major components are indicated: **(1)** drive system (servo motor and odometer), **(2)** adjustable aluminum framing (80/20), **(3)** roller chain with attachment points, **(4)** sorbent packet, **(5)** tensioning arm, **(6)** basin, **(7)** pH probe and **(8)** pump.

The system's belt was designed to accommodate AER-based sorbents in a variety of commercially available forms including powders, beads, sheets, fibers and membranes, or novel yet to be developed forms such as fabrics or foams. Larger sorbent structures such as sheets, membranes, fibers, fabrics, and foams can be directly attached to the drive chains using a fastener, whereas smaller powders and beads must first be contained within mesh pouches that are in turn attached to the drive chains using fasteners. Mesh bags are commonly used for containing AERs when exposed to water and soil for environmental studies and water treatment.[31–34] The drive chain is attached to a drive motor that rotates the belt either in a continuous or stepwise fashion.

As most AERs are quite hydrophilic,[35] if immersed in liquid water they quickly absorb substantial quantities of water and rapidly (< 30 min) release $CO_2$ into solution but take longer to dry and load $CO_2$ (2–4 h) in ambient air. The time to refill with $CO_2$ is directly correlated with ambient air wind speed, temperature and initial sorbent water loading and inversely correlated with the air's relative humidity (RH). To increase the productivity of the system, the belt rotation speed can be altered based on ambient conditions with sufficient knowledge of the sorbent's drying and $CO_2$ loading time as a function of wind speed, temperature and/or RH. To adjust the relative time of immersion and exposure to wind, the sorbent belt and drive motor are attached to a support frame, where the travel distance of the belt in air is longer than that in the liquid. Depending on the desired ratio of the belt exposed to air and to liquid and the flexibility of the sorbent forms, the belt can be rotated in a single large loop (**Figure S1 and S2**), woven back and forth to reduce the physical space and areal footprint needed to support the belt when exposed to air, or split into multiple concurrently or independently moving loops (**Figure 3**).[36]

1 g $CO_2$ per day (*1g*) system design. To demonstrate the $CO_2$ capture concept under controlled wind and laboratory conditions, we first designed a sorbent conveyance system (**Figure 2**) to circulate commercially available AERs—A501 sorbent beads (Purolite) contained within mesh packets or Excellion (SnowPure) sheets—within a laminar flow hood. Both of these sorbents are polystyrene-based AERs with quaternary ammonium functional groups manufactured for water treatment and were found to be biocompatible with cyanobacteria (see section below on sorbent preparation). The system was designed to hold ~16 g of sorbent that we estimated could deliver ~1 g of $CO_2$ per day, so we called this the *1g* system. The *1g* system includes: **(1)** a drive system with a motor (ServoCity) and odometer to track the total belt movement, **(2)** aluminum 80/20 T-slot framing to support the belt with axle mounting locations capable of accommodating different ratios of sorbent belt exposed to air or liquid, **(3)** a roller chain to connect **(4)** the sorbent packets to create the sorbent belt, **(5)** a tensioning arm to assist with belt alignment and keep the chain attached to the drive sprocket, **(6)** a basin to contain aqueous medium to capture the delivered $CO_2$ (either abiotic or for cultivating cyanobacteria), **(7)** a pH probe (Fisher Scientific) to monitor $CO_2$ delivery and removal, and **(8)** a peristaltic pump (Longer Pump) for circulating medium in the basin. All components sitting at or below the water level were made from biocompatible materials to avoid interference with cyanobacterial or microalgal cultivation, including nonstandard plastic versions of common components (axle mounts, pillow blocks / bearings, stand offs, chain, bolts and screws). In cases where substitutions were not available, in-house 3D printers and machining tools were used to create biocompatible custom hardware.

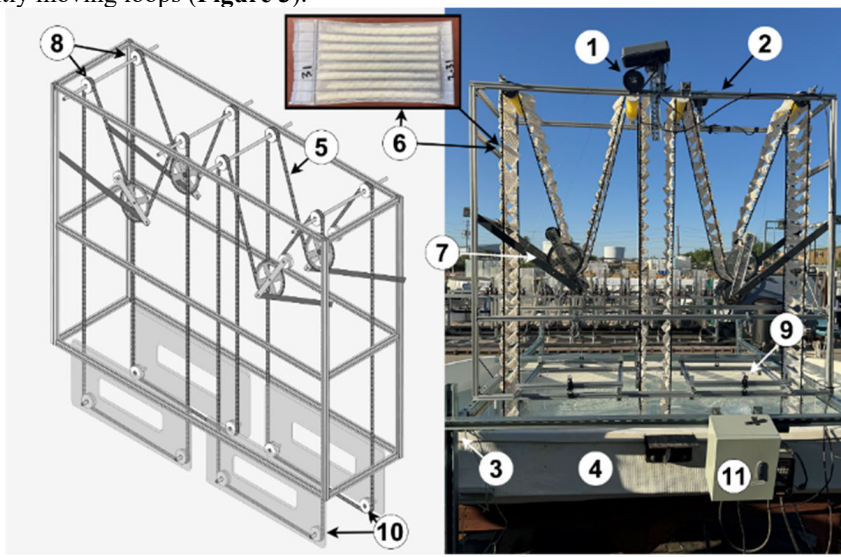

**Figure 3.** Schematic (left) and photo (right) of the *100g* system. All major components are indicated: **(1)** drive system (servo motor, control system, odometer), **(2)** aluminum superstructure frame (80/20 T-slot), **(3)** a steel Unistrut base, **(4)** fiberglass pond, **(5)** plastic roller chain, **(6)** double-wide packet sorbent belt, **(7)** heavy independent tensioning arms, **(8)** chain guides, **(9)** lifting system, **(10)** static sprockets and **(11)** automation control box.

100 g $CO_2$ per day (*100g*) system design: Based on learnings from the *1g* system, the design was scaled up to deliver 100 g $CO_2$ per day and thus called the *100g* system. It was designed to evaluate the performance of the system under ambient outdoor conditions to identify issues that must be addressed to further scale up the system. The major components of the *100g* system included **(1)** a more complex drive system that included a feedback control system to adjust the belt speed based on ambient wind speed using an adaptive algorithm in addition to the servo motor and odometer, **(2)** an adjustable aluminum superstructure frame (80/20 T-slot) to support the system in combination with **(3)** a steel Unistrut base to span the pond without contacting the **(4)** fiberglass pond, **(5)** a much longer plastic roller chain spaced further apart to attach **(6)** a wider sorbent belt that accommodated two connected sorbent packets (i.e., a double-wide packet), **(7)** heavy-weighted independent tensioning arms to maintain tension and prevent flapping and uncontrolled motion driven by wind forces, **(8)** custom-designed chain guides to keep the chain connected to the sprocket in the event of slippage or wind-driven chain movement, **(9)** a lifting system to pull all system components above the water as necessary for cleaning and maintenance, **(10)** static sprockets mounted below the water line, and **(11)** an automation control box. As with the *1g* system, all components in the *100g* system sitting at or below the water level were biocompatible to support cultivation studies with microalgae or cyanobacteria. The chains of the sorbent belt were attached to co-axially mounted sprockets connected by a shaft including one driven by a single servo motor to precisely control the belt speed. In order to increase the number of packets exposed to air within the physical footprint of the straight path of one side of the raceway pond, the top segment of the belt path was extended downward and back up in a V shape (Figure 3).

Sorbent preparation and ion exchange: A501 (Purolite) AER, with an ion-exchange capacity of ~2 milliequivalents (meq) per gram (dry basis), was purchased in the chloride form, which is not active for moisture-driven DAC. The desired quantity of sorbent was added to a fresh solution of 1.0 M $NaHCO_3$ (baking soda) and soaked overnight while gently stirring this process was repeated once, and then the sorbent was rinsed multiple times in deionized (DI) water to remove excess $NaHCO_3$. Lower concentrations of $NaHCO_3$ are also suitable as long as they have at least a 20−30-fold molar excess over the number of charge sites in the AER to be exchanged (e.g., 1 g sorbent in ≥ 100 mL of 0.5 M $NaHCO_3$). A similar process was used to exchange Excellion (SnowPure) AER sheets, which has well established moisture swing DAC properties.[5,20]

Sorbent containment: Food-grade nylon woven mesh sheets with 25 μm pores (LBA) were first cut into 18 x 10 cm sheets. Two sheets were then thermally sealed to form eight elongated tubes with 0.75 cm diameter either manually with an impulse sealer (American International Electric) or using a custom-built pneumatic thermal press with parallel wires that are heated with an electric current to create a seam to contain the sorbent beads (**Figure S3A**). The packets were then sealed on one end with an impulse sealer, filled with ~1 or 2 g of $NaHCO_3$-exchanged A501 sorbent and after being filled, sealed on the other end with an impulse sealer (**Figure 3 item 6**). A secondary seal was added on each end using an impulse sealer to prevent sorbent spills due to mesh wear. Rivets were then added to one end of each packet to snap two packets together to make a double-wide packet and were attached to two parallel chains using nylon nuts and bolts to form a sorbent belt (**Figure S3B**) that could be continuously cycled between air and a liquid alkaline medium. Several drawings and additional details are provided in the Taylor et al. patent application.[30]

Sorbent water uptake: The water uptake of an Excellion sheet, A501 beads loaded in a nylon mesh packet, and an empty nylon mesh packet were measured using a load cell (FUTEK; **Figure S4A**) by first soaking samples for 10 min in deionized water, then hanging the wet samples from the load cell for 150 min in ambient lab conditions (~22 °C, ~40% RH). The percent water uptake was plotted by dividing the measured weight by the initial dry mass of the sample (**Figure S4B**).

Lab-scale sorbent packet $CO_2$ binding capacity and delivery evaluation: Mesh packets were loaded with either 1 or 2 g of AER sorbent beads or AER sheets, dried overnight in ambient air (~22 °C, ~15% RH) and submerged in 1 L of 10 mM $Na_2CO_3$. Then, the pH of the solution was recorded every 30 s for 1 h with constant stirring using a magnetic stirrer at 100 rpm. The pH change was used to calculate the amount of $CO_2$ delivered by the sorbent using calculations described by Shesh et al.[37] A similar procedure was used to measure $CO_2$ delivery from Excellion AER sheets.

Sorbent drying and $CO_2$ loading evaluation in a wind tunnel: About 10 g of A501 sorbent was loaded in a polyester mesh bag or dispersed into ten mesh tube packets each with ~1 g of sorbent and then mounted parallel to the wind flow direction within a 3D-printed holding frame sized to fill the 4 in x 4 in (101.6 mm x 101.6 mm) cross section of a closed-loop wind tunnel (volume ~560 L); the wind tunnel is described in section 3.4 of Barbosa's thesis.[38] The sorbent mesh packets or mesh bag were exposed to ambient air overnight to equilibrate with atmospheric $CO_2$ and $H_2O$ vapor partial pressures before being submerged in an aqueous solution of 0.1 M $Na_2CO_3$ for 1 h to desorb the $CO_2$. The now $CO_2$-depleted sorbent mesh bag or packets were then removed from the $Na_2CO_3$ solution, rinsed with DI water and left for five min on the sample holder to remove excess water from the surface, so it would not get blown into the wind tunnel. The sorbent packet frame was then inserted in the wind tunnel chamber at 22 °C with the wind velocity set to the desired wind speed (0.7, 1.6, 2.5 or 5 m/s) and moisture levels were maintained between 3 and 5 parts per thousand (ppt) $H_2O$ or about ~15% relative humidity (RH) using a condenser to actively remove water from the air stream to maintain this moisture level.

Pond-scale $CO_2$ capture and delivery evaluation: A 4.2 $m^2$ pond (Commercial Algae Professionals), equipped with a YSI 5200A-DC (YSI Inc., Yellow Springs, OH, USA) water quality monitoring system that simultaneously measures pH, pond water temperature (°C), dissolved oxygen saturation (%), and salinity (g/L) as described by McGowen et al.,[39] was retrofitted to replace the paddlewheel with a centrifugal pump to minimize passive $CO_2$ capture independent of the DAC system. The pond was filled to a 20-cm depth (~840 L) with 17.5 mM $Na_2CO_3$ (pH ~11) to provide capacity to store the $CO_2$ delivered by the DAC system, which reduces the pH and pushes the equilibrium from $CO_3^{2-}$ toward $HCO_3^-$. The DAC system belt motors were turned on and run at two speeds, such that packets completed a 4-h wet/dry cycle during the night when sorbent drying and loading are slower due to lower temperatures and wind speed and higher RH, and a 2-h cycle during the day when sorbent drying is faster due to higher temperatures and wind speeds and lower RH. When the pH of the pond medium was reduced to 9.75 or below, 10 M NaOH was added to bring the pH back to ~10.25 and thereby convert some of the $HCO_3^-$ to $CO_3^{2-}$ to increase the medium's capacity to bind additional $CO_2$ delivered by the DAC system.

Technoeconomic and life cycle analysis (TEA/LCA): The commercial feasibility and environmental impacts of a moisture-driven DAC system were evaluated through concurrent techno-economic analysis (TEA) and life cycle analysis (LCA). This analysis covered the costs and greenhouse gas emissions associated with an $n^{th}$-of-a-kind facility based on current system performance to capture $CO_2$ and deliver it into an alkaline brine with additional costs and emissions to extract $CO_2$ from brine, purify and compress it to 15 MPa for use, conversion or storage such that it could be compared with other thermally-driven DAC systems. First, an engineering process model of the system was developed that captures mass and energy flows. The model is depicted by the process flow diagram in **Figure 4**.

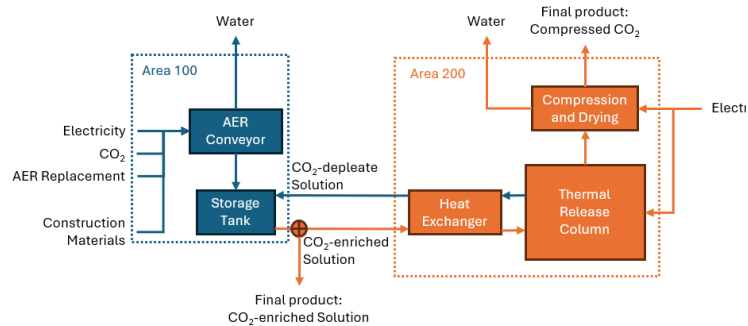

In Area 100, AER was cycled between the air and an alkaline aqueous solution by an overhead zig-zag conveyor transferring $CO_2$ to the solution that is inspired by designs from Lackner.[40] This $CO_2$-enriched solution is suitable for cultivating microalgae and forming stable carbonate minerals. In Area 200, the solution is heated with an immersion heater to release the $CO_2$ in a thermal release column; heat exchangers preheat the solution while recovering heat from the hot solution leaving the column. Key assumptions and inputs used to calculate the mass and energy flows are described in the supplemental information. Input variables used in the calculations are listed in **Table S1**, **Table S2** and **Table S3**. Additional assumptions for carbon intensities, material and energy price data, equipment cost data, life cycle inventory, capital investment and operational expenditures are listed in **Tables S4, S5, S6, S7, S8, and S9,** respectively. These parameters form the basis of two scenarios, a practical scenario representative of experimental results to date and an aspirational scenario representative of improved performance by future development.

**Figure 4.** Process flow diagram of the moisture-driven DAC system with thermal $CO_2$ release and compression. AER: anion exchange resin.

LCA was performed by integrating energy and mass data from the foundational process model with life cycle inventory data. As the objective of carbon capture is frequently greenhouse gas reduction, the results of LCA were combined with the results of TEA and presented as a single metric, the minimum selling price (MSP) of a net tonne of $CO_2$ captured. A net tonne of $CO_2$ captured is defined as the difference between $CO_2$ removed from air and $CO_2$ released through system operation, such as in electricity generation. TEA was performed using a 30-year discounted cash flow rate of return (DCFROR) model to determine the MSP. Commercial feasibility is often assessed by the MSP, or the minimum price at which a product can be sold to achieve a net present value of zero dollars over the system's lifetime while accounting for a reasonable return on investment. LCA utilized the International Organization for Standardization (ISO) standards ISO 14040 and ISO 14044 to provide the framework for conducting the assessment, ensuring methodological rigor and consistency. The 100-year global warming potential values from the International Panel on Climate Change (IPCC) AR6 were used[41]—1 for $CO_2$, 28.9 for $CH_4$, and 273 for $N_2O$. TEA and LCA were performed for two scenarios, practical and aspirational, which were defined based on current experimental system performance and targeted future performance.

A global sensitivity analysis was conducted based on the methods of Morris[42] to identify important areas to focus further research effort. For the global sensitivity analysis 10 random and unique designs were selected by varying the values of each variable. The elementary effect of each variable in each design was calculated as defined in Morris.[42] The mean and standard deviations of these elementary effects of a variable across all designs were recorded.

Also, a Monte Carlo simulation was performed, which is a method of defining the impact of uncertainty in input variables on MSP. Monte Carlo simulation uses random sampling, running the model with inputs selected from a pre-defined distribution, to predict outcomes based on probability distributions. We used uniform distributions for the variables not experimentally defined in the Monte Carlo simulation while the other variables were held constant.

## 2. Results and Discussion

Sorbent selection and mesh packet design. The most critical material in the DAC system is the sorbent. A501 AER beads (~420–1200 μm diameter) were selected due to their microporosity to increase gas and liquid transport throughout the AER particles during $CO_2$ capture and release, respectively, and for their stability in the alkaline medium (pH 9–11) needed to retain and concentrate inorganic carbon (IC) in the medium. We developed a method for containing the sorbent beads within thin elongated tubes made of nylon mesh with 25 μm pores. The tube packets are shown in **Figure 2** and **Figure S3B**. A 0.75 cm tube diameter was found to be a good compromise between a practical width sufficiently large to facilitate filling the tubes manually with sorbent beads versus smaller diameters that increase the surface area to volume ratio. We did not find a significant difference in sorbent loading rates when filling the packets with 1 versus 2 g of sorbent or when leaving empty tubes between filled tubes. Filling the tubes with 2 g of sorbent still provided ~30% empty volume per tube for the sorbent to swell during wet-dry cycles.

Due to its high charge content, A501 contained in a hydrophilic mesh packet absorbs ~330 wt.% water (**Figure S4B**), which is beneficial for rapidly releasing captured $CO_2$, but it slows sorbent drying before $CO_2$ can effectively bind, and also increases the water usage and environmental impact. Sorbent drying and $CO_2$ loading when contained in a mesh packet with ten elongate tubes compared to a mesh bag with a single compartment were measured in a closed-loop wind tunnel, in which dry air (15% RH) at 25 °C was circulated at 5 m/s. As shown in **Figure 5**, the mesh tube packets required 30 minutes of drying before $CO_2$ began to bind and 80 minutes to load to 90% of the sorbent's $CO_2$ capacity, whereas the mesh bag took about 50 minutes of drying to being loading $CO_2$ and about 275 minutes to load to 90% of the sorbent's $CO_2$ capacity. **Figure S5A** shows this maximum uptake rate of the mesh tube packets was 4.3-fold faster than when the same sorbent beads were contained within a single mesh bag, which allows the sorbent beads to clump together (**Figure S6**); attempts to physically disperse large clumps of the sorbent particles in the mesh bag was inconsistent and is the likely source of large variation in drying and $CO_2$ uptake rates observed over multiple runs. The small reduction (70 μmol $CO_2$ per g sorbent) in apparent $CO_2$ capacity of the sorbent when contained in the mesh bag (**Figure S5B**) is likely due to leakage of $CO_2$ from ambient air into the wind tunnel air chamber, which can underestimate the sorbent capacity of experiments with extended time below 300 ppm. Time-dependent sorbent $CO_2$ release was measured by immersing the dry loaded sorbent packet into a 1 L solution of 10 mM $Na_2CO_3$. **Figure 6** shows that ~90% of $CO_2$ sorbed onto the sample was delivered within 15 min, which is 3–4 times faster than an AER sheet of Excellion SnowPure. We hypothesize the structural materials used to form the Excellion

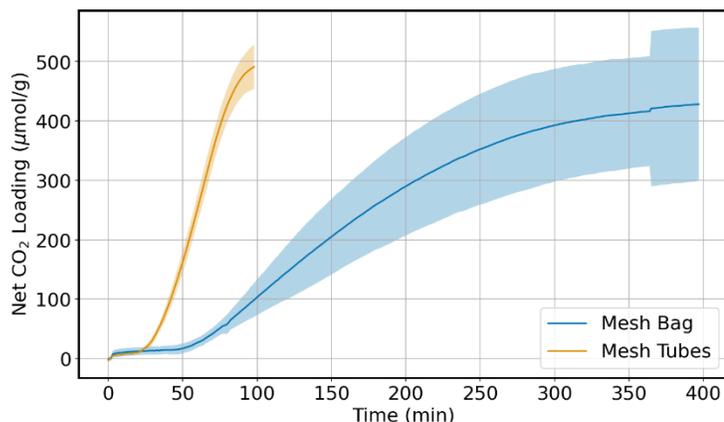

**Figure 5.** Sorbent drying and $CO_2$ loading from air (5 m/s, 15% relative humidity, 25 °C) comparing a mesh tube vs mesh bag form factor. Data areshown as an average of three (tube) or four (bag) replicates with a colored area showing a 95% confidence interval.

sheet likely slowed $CO_2$ transport and release into the medium.

X. Wang et al. evaluated $CO_2$ loading and unloading of several related AER beads that were ground into particles of varying sizes. The largest particles used in that study were isolated with an 18–50 mesh sieve that corresponds to particles with diameters between 0.3–1.0 mm that is comparable to the A501 beads (~0.6 mm) used in the present study.[43] X. Wang's data for IRA-900 (DuPont), which has a similar chemical composition, bead diameter, and microporous morphology as A501, showed much slower desorption (90 min) that we attribute to using 99.9% RH air to release $CO_2$ rather than liquid used in our study, but had much faster $CO_2$ absorption (30 min) since presumably the resin bound much less water from exposure to humid air than from liquid immersion as in our study. Thus, the overall time to run a complete cycle for loading and unloading $CO_2$ were both about 2 h regardless of whether liquid water or humid air were used for desorption data. X. Wang's study also reported results for another microporous anion exchange resin D201 (Comcess) ground and isolated with the same size mesh that also took around 30 minutes to load $CO_2$; however, it unloaded $CO_2$ in 30 minutes, which is about twice as fast as IRA-900.[43] Song et al. reported even faster (5–6 min) $CO_2$ loading times for D201 when it was ground into a fine powder (400 mesh, ~40 μm diameter) to increase surface area and when exchanged with phosphate.[44] T. Wang showed $CO_2$ sorption rates can also be further improved by increasing porosity.[45]

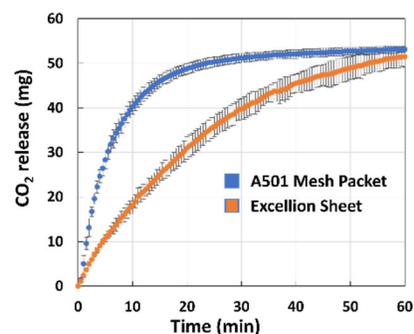

**Figure 6.** $CO_2$ released into 10 mM $Na_2CO_3$ by 1 g of A501 sorbent beads within a mesh packet compared to a 20.3 cm x 1.7 cm (~1.76 g) AER sheet of Excellion.

<u>1 g $CO_2$ per day (*1g*) system design and evaluation</u>. To demonstrate the $CO_2$ capture concept under controlled wind and laboratory conditions, we first designed a sorbent conveyance system (**Figure 2**) to circulate the mesh packets with A501 sorbent beads and Excellion sheets in a laminar flow hood. The system was designed with a sorbent belt composed of 16 mesh packets each filled with 1 g of A501 sorbent, that we estimate would deliver ~1 g $CO_2$ per day so we called this the *1g* system. The *1g* system was run with a 4-h cycle (1 h in 20 mM $K_2CO_3$, 3 h exposed to dry ambient air) with and without an A501 sorbent belt. As shown in **Figure 7**, the pH of the 12-L 20 mM $K_2CO_3$ solution decreased more rapidly when the sorbent belt was attached than through passive diffusion without the belt, indicating additional $CO_2$ is being delivered by the sorbent packets on the belt. The pH change converts to ~7 g of $CO_2$ delivered over 44 h with the A501 belt, compared to 4.6 g without the belt, which is 1.3 g $CO_2$/day of net $CO_2$ delivered by the *1g* system. This is substantially lower than the ~4 g $CO_2$/day that was predicted from the measured capacity of fully dried individual sorbent packets (Figure 6). We hypothesize this was primarily due to inadequate packet drying and $CO_2$ loading within the 3-h timeframe when a packet was in the air due to slow air flow in the laminar flow hood (~0.4 m/s), but also due to less $CO_2$ being delivered by the sorbent into the $K_2CO_3$ solution each cycle as it slowly fills with $CO_2$ and $CO_3^{2-}$ in solution is converted to $HCO_3^-$.

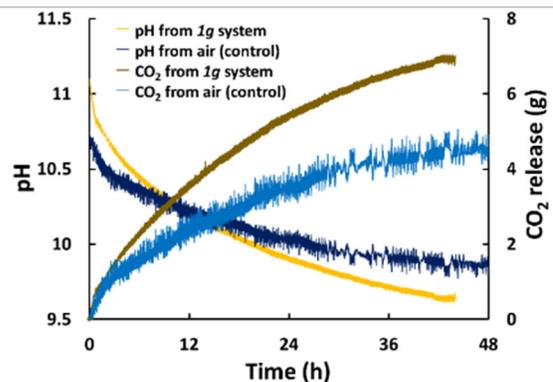

**Figure 7.** $CO_2$ delivered into a 20 mM $K_2CO_3$ solution in the updated 1g system with (orange) or without (blue) 16 g of A501 pixie stick packets. The descending lines correspond to pH and the ascending lines to $CO_2$ release.

When the sorbent is immersed in a solution, it will unload $CO_2$ only until the $HCO_3^-$ sites on the sorbent reach equilibrium with those in solution, since some of the $CO_3^{2-}$ produced on the resin when $CO_2$ is released (see Figure 1) will subsequently be exchanged with $HCO_3^-$ from solution until the concentration of $HCO_3^-$ on the sorbent and solution reach equilibrium. In this case, 12-L of 20 mM $K_2CO_3$ has a capacity to hold no more than ~10.5 g $CO_2$ when all $CO_3^{2-}$ has been converted to $HCO_3^-$. Thus, after ~7 g of $CO_2$ is accumulated in the buffer (Figure 7), it contains ~13 mM $HCO_3^-$, which will prevent a substantial fraction of the $CO_2$ captured on the sorbent as $HCO_3^-$ from being released into solution. This underscores the need for periodically extracting $CO_2$ from the buffer to maximize the fraction of the sorbent's captured $CO_2$ that is delivered into solution..

The *1g* system was also evaluated with 40 Excellion AER sheets (~20.3 cm long x 1.7 cm wide) attached to a chain-driven belt as shown in **Figure S7** and the basin was filled with ~12 L of 1 mM $K_2CO_3$ to evaluate a lower capacity $CO_2$ storage medium. The system was evaluated with three different cycle durations lasting 1.5, 2 or 4 h using the same 3:1 ratio of time that the belt was exposed to air versus the alkaline aqueous medium. As shown in **Figure S8A,** the pH of the system with the Excellion sorbent dropped much more rapidly than the control without sorbent until the solution became saturated. After saturation at pH ~8.2, the sorbent-delivered $CO_2$ reached an equilibrium with $CO_2$ outgassing to ambient air. A moisture meter showed the Excellion strips had measurable moisture on the surface at the end of the air-contacting phase just prior to being submerged during the 2-h cycle but not during the 4 h cycle. Interestingly, the amount of $CO_2$ delivered during each cycle did not change with the time to complete a wet-dry cycle; we hypothesize that slower cycles allowed for higher $CO_2$ loading so that more $CO_2$ was delivered when the sorbent contacted the medium, but that was offset by the belt rotating slower to have fewer cycles per day for $CO_2$ delivery. The equilibrium pH of the medium was somewhat higher for the 2- and 4-h cycles than for the 1.5-h cycle, which we hypothesize was because the sorbent loaded more fully and thus had a higher driving force to release $CO_2$ into the medium compared to the 1.5 h cycle. In all cases the Excellion belt delivered $CO_2$ much faster than the background uptake from ambient air and the sorbent pushed the equilibrium pH below what could be obtained without a sorbent for all but the 1.5 h cycle. These results also indicate that the air flow provided by the hood (~0.4 m/s) is limiting the amount of $CO_2$ delivered per cycle.

100 g $CO_2$ per day (*100g*) system evaluation: Based on learnings from the *1g* system, the design was scaled up to deliver 100 g $CO_2$ per day and thus called the *100g* system (**Figure 3**). It was designed to evaluate the performance of the system under ambient outdoor conditions to identify issues that must be addressed to further scale up the system. The *100g* system support structure was assembled with a rectangular geometry to maintain a 9:1 ratio of packets exposed to air vs immersed in the $Na_2HCO_3$ medium based on laboratory $CO_2$ delivery experiments showing $CO_2$ was delivered within 20–30 min (Figure 6) and the packets dried and loaded with $CO_2$ within 2–4 h (Figure S5). The 3:1 ratio from the original *1g* system had inadequate time for drying (or excess time for $CO_2$ delivery). Initial trials with the *100g* system used a belt with an inner and outer loop (Figure S1 and S2), which came off the track due to an excessive angle between the plastic chain and sprockets, differences in the rotating speeds between the drive chain and the freewheeling chain, and insufficient tension. A second-generation system (Figure 2) was then designed and built that created two parallel loops with a wider turning radius and independent tensioning arms to enable smoother operation. Additionally, the use of two independent drive motors provided redundancy should one system fail. This design also increased the number of packets that could be installed from ~256 to 376, which increased the theoretical $CO_2$ capacity of the belt from ~10 g $CO_2$ per full wet/dry cycle by 50% up to 15 g $CO_2$ per full wet/dry cycle.

In April 2024, the *100g* system was attached to a 4.2 m$^2$ (areal surface area; ~840 L) raceway pond at the Arizona Center for Algae Technology and Innovation (AzCATI) in Mesa, AZ (Figure 2). The system was initially operated with fixed 2-h wet/dry cycles during the day (when light was detected by an onboard photosensor). During daylight hours wind speeds and temperatures are typically higher and relative humidities lower than at night. This results in more ideal drying and $CO_2$ capture conditions during the day. At night, the system was slowed to 4-h wet/dry cycles due to slower drying and $CO_2$ loading.

**Figure 8** shows the inorganic carbon (IC) from the pond with the *100g* system after subtracting background $CO_2$ uptake from air (see also **Figure S9**) as measured by the IC in the control pond without the *100g* system during an outdoor trial from May 1 to May 8, 2024.[36] The average $CO_2$ delivered each day increased during the trial up to around 100 g $CO_2$ per day, which is what the *100g* system was designed for. A model was developed to estimate the amount of $CO_2$ delivered for the time the sorbent was exposed to a particular wind speed based on data collected in the wind tunnel (**Figure S5**) and **Equations S17, S18 and S19** and empirical curve fits (**Figure S11** and **Figure S12**). The model was used to estimate the amount of $CO_2$ delivered based on wind speeds recorded by a nearby weather station and the amount of time the sorbent packets were exposed to that wind speed throughout the seven-day trial. The model predicted 762 g $CO_2$ should have been delivered during the trial, which is within 3% of the measured value of 782 g. It is worth noting that during this period the average temperature and RH (24.4 °C and 16.7% RH) were quite similar to the conditions when wind tunnel data were collected (22 °C and 15% RH) and used to build the $CO_2$ loading model, which likely explains the model's good fit (i.e., the effects of temperature or RH on $CO_2$ loading are absent).

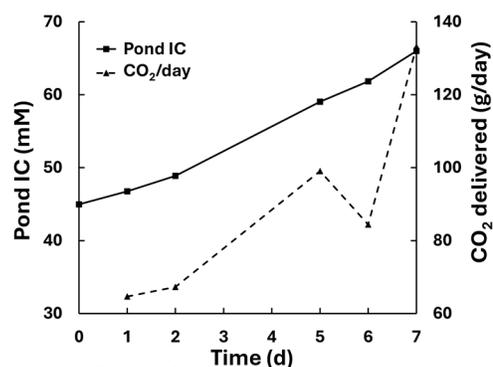

**Figure 8.** Pond inorganic carbon level and $CO_2$ delivered each day from DAC (after subtracting background $CO_2$ uptake from air).

Rittmann, Lackner, Flory et al. also developed and reported a moisture-driven DAC system that used liquid water immersion to release $CO_2$.[29] In that study, the commercial Excellion AER sorbent sheets maintained their integrity, and their performance could be restored by washing with baking soda after months of wet/dry cycling and exposure to harsh outdoor conditions in Arizona. The approach of Rittmann et al. and the present study both benefit from using moisture to release $CO_2$ from the sorbent without the need for external heat or vacuum pressure energy. However, the Rittmann et al. approach had several limitations that were overcome in the current approach including, **1)** $CO_2$ release into liquid water by Excellion AER sheets was 3–4 times slower than that by the sorbent beads used in this study, **2)** $CO_2$ release into liquid water led to a reduced $CO_2$ capacity from $Cl^-$ accumulation in the water immersion tank, although it could be restored by washing in baking soda, whereas the current approach uses alkaline medium enriched with 35 mM $HCO_3^-$ to minimize the influence of non-moisture-swing-active anions in the medium such as $Cl^-$, **3)** the low $CO_2$ capacity of water used to release $CO_2$ by Rittmann et al. led to an inefficient and costly scheme to store, extract and deliver $CO_2$ whereas the current approach uses an alkaline medium to retain the captured $CO_2$ until it is used without loss to the environment, **4)** the air collector design was sensitive to wind direction whereas the current system has been designed to capture $CO_2$ passively from natural wind flows from any direction, and **5)** the Rittmann et al. system was operated in a batch process that may be difficult to scale,[29] whereas the current system operates in a continuous process. Both studies identified sorbent capacity, stability, drying rate and tolerance toward salts in the water source as key parameters to improve in order to enable commercial demonstration.

Commercial feasibility and environmental impacts. The costs and greenhouse gas emissions associated with $CO_2$ capture were assessed for an $n^{th}$-of-a-kind facility operating on this novel system and compared with other thermally-driven DAC systems. The costs are presented in **Figure 9** for a practical scenario that assigned specific values to performance parameters that are representative of experimental results to date as well as those that include specific improvements: A) increasing sorbent lifetime, B) increasing sorbent capacity, C) decreasing the drying and $CO_2$ loading time, and D) reducing the AER price. The aspirational scenario includes improvements A, B and C, but excludes D since AER price is not directly addressed in this work. The best-case scenario specifically assigned the minimum value of all input data. Figure 9 also distinguishes between two different end products: 1) A final product of a $CO_2$-enriched solution suitable for microalgae cultivation and mineralization, which is based on experimental results demonstrated in this work, is designated as 'solution'; 2) a final product of $CO_2$ compressed to 15 MPa suitable for geologic sequestration, which models costs to extract, purify and compress $CO_2$ from the $CO_2$-enriched solution, is designed '15 MPa'. Cost and energy consumption are presented for one net tonne of $CO_2$ captured into solution or further compressed to 15 MPa, where $CO_2$ net captured is defined as the difference between $CO_2$ removed from air and $CO_2$ released from the activities involved in capture, such as burning natural gas. The mean values of all input data ranges were used unless specific values were assigned by a scenario.

Large cost drivers in the practical scenario are the energy used by the conveyor and the price of AER, which are greatly reduced in the aspirational scenario. The initial investment in AER and the cost of AER replacement every five years contribute $200 per tonne to net capture cost. The energy used by the conveyor is important, not only because of the electricity price, but because of the cost of motors and the penalty of indirect $CO_2$ emissions associated with electricity generation. Almost 200 kg of indirect $CO_2$ emissions are released to generate the electricity used by the system in capturing 1 tonne of $CO_2$, increasing the size of the system and therefore the total costs by 25% (dark purple bar in Figure 9). The aspirational scenario was defined with feedback from sensitivity analysis and is elaborated on at the end of this section.

The practical scenario costs $950/tonne $CO_2$ net removed and uses 490 kWh/tonne $CO_2$ net. The major cost drivers in the practical scenario are the water uptake of the sorbent—where higher values can cause exceeding the carrying capacity of the conveyor—the carrying capacity of the conveyor, sorbent capacity, and $CO_2$ loading rates, which are

also impacted by water uptake and drying time. Low adsorption rates lead to higher energy usage of the conveyor system because a decrease in adsorption rate increases the belt length and thus increases friction between the belt and its track. While increasing water uptake tends to have an adverse effect on adsorption rates, it also directly impacts the length of the conveyor. Given the conveyor has a limited carrying capacity of 167–278 kg per meter, carrying more water out of the tank increases the length of conveyor required to stay within the carrying capacity.

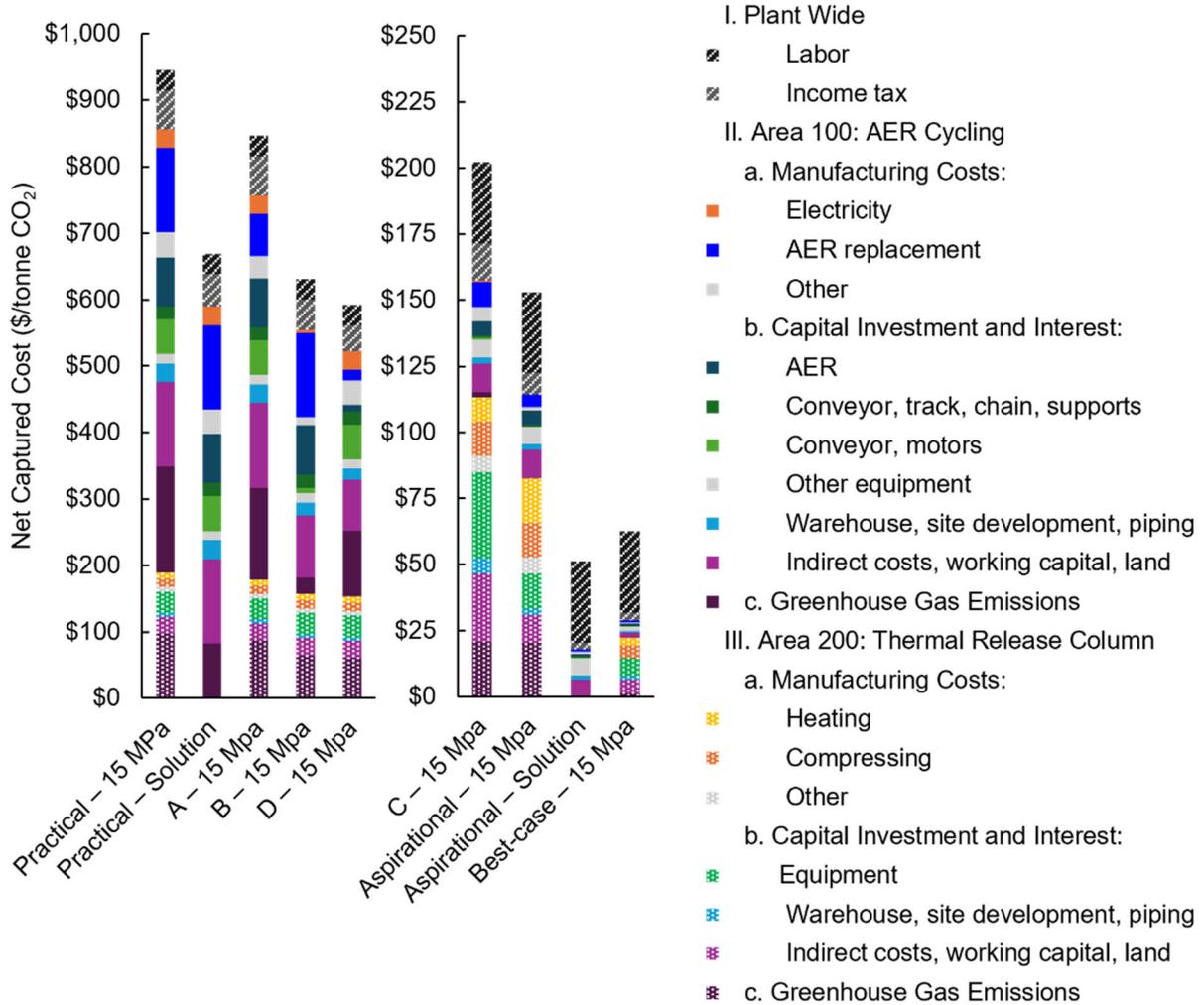

**Figure 9**. The cost to capture a net tonne of $CO_2$ (i.e., minimum selling price) delivered into solution (solution) or thermally extracted from solution and compressed to 15 MPa (15 MPa) for the practical scenario based on data from this study as well as several individual improvements to A) increase AER lifetime, B) increase AER capacity, C) decrease resin price to 7 $/kg, D) increase AER loading rate with quicker drying, decreased water uptake, and increased desorption rate. The aspirational scenario combines improvements A, B and C, and the best-case scenario assigns the minimum value of all input data. Each expense is shaded in proportion to its relative contribution to the total cost.

An aspirational scenario improves several key parameters to values we think are feasible with additional research to bring the costs and energy use down to $160/tonne net captured and 350 kWh/tonne net captured with a best-case scenario of $66/tonne net captured. A scenario that captures atmospheric $CO_2$ into aqueous medium Additionally, in the aspirational scenario, the optimal heat exchanger effectiveness is lower increasing energy use but decreasing heat exchanger capital costs (Figure 9). **Figure 9** also shows cost reductions from specific improvements to the practical scenario when done independently, including, A) increasing sorbent lifetime, B) increasing the sorbent capacity from 0.75 to 4.0 mmol $CO_2$/g sorbent, C) reducing water uptake of the sorbent to 50 wt.% by increasing hydrophobicity, which should correspondingly reduce the sorbents drying and $CO_2$ loading time we estimate from 2 hours to 1 hour, and D) reducing the AER price to 7 $/kg. Wang et al. reported AER sorbents with ~3.4 mmol/g capacities although only ~1 mmol/g was achieved within 1 hour of loading.[46] Comparatively, liquid sorbent technology achieves capture and compression with a similar projected cost between $150–170/tonne net captured, but has a much higher an energy use of 1,700–2,400 kWh/tonne net captured.[47,48] Meanwhile the lower operating temperatures of solid sorbent technologies allow for the utilization of a heat pump and electricity for heating. Energy use can be as low as 1,300 kWh/tonne net captured on a clean grid like in upstate New York but can get as high as 6,000 kWh/tonne net captured depending on the carbon intensity of the grid with cost ranging from $230/tonne net captured on a clean grid to $1,200/tonne net captured on higher carbon intensity grid.[49]

If the $CO_2$-enriched solution is treated as the final product, the cost of capture decreases by $280 per tonne in the practical scenario and $109 in the aspirational scenario. The $CO_2$-enriched solution is suitable for cultivating microalgae or forming stable carbonate minerals. The exact end-use of the $CO_2$ will determine if it is permanently sequestered. Frequently microalgae is used to generate biofuels or as a food product; the $CO_2$ in those end-use scenarios would not be captured and sequestered, but this technology would be a renewable source of $CO_2$ regardless.

Monte Carlo simulation helps shed light on the probability of achieving the best-case scenario. The TEA presented in Figure 9 employs deterministic scenario modeling, wherein specific values are assigned to model inputs to represent distinct operational conditions. In the practical and aspirational scenarios, these values typically correspond to the mean of each input's defined range, unless otherwise specified by empirical performance data provided in **Table S1**. The best-case scenario, by contrast, utilizes the minimum values within each input range to reflect an optimistically favorable configuration. While these single-point scenarios offer insight into potential system performance under fixed assumptions, they do not account for variability or uncertainty in input parameters. To address this limitation, Monte Carlo simulation is employed as a probabilistic modeling approach. This method involves repeated random sampling of input parameters from uniform distributions defined by their respective ranges. By executing the model across a large number of iterations, the Monte Carlo simulation generates a distribution of possible outcomes, thereby enabling the quantification of uncertainty and the estimation of outcome probabilities,

The results from the Monte Carlo simulation of the minimum selling price (MSP) of captured and compressed $CO_2$ to pipeline specifications (15 MPa) with this moisture-driven DAC process are depicted in **Figure 10**. Based on the current performance of the as-built system, with additional cost to thermally extract $CO_2$ from the aqueous brine, the projected cost is between $530–1,500/net tonne $CO_2$ captured with 95% confidence, with a median cost of $930/tonne. The aspirational scenario reduces cost substantially to $120–260/net tonne $CO_2$ captured with 95% confidence, with a median cost of $170/tonne. The best-case scenario falls outside the 95% confidence interval. Achieving the aspirational scenario greatly increases the likelihood of the system being cost-competitive with industry leading DAC technologies.

The order of magnitude range in the practical scenario is a result of increased sensitivity to many variables in the practical scenario. The combination of high water uptake and low effective $CO_2$ binding capacity in the practical scenario led to high energy usage by the conveyor system and a lower $CO_2$ removal efficiency. This makes the practical scenario particularly sensitive to the carrying capacity of the conveyor and the amount on non-resin material affixing the resin to the conveyor, variables that the model is not generally as sensitive to. No systematic assessment of the cause of uncertainty was conducted, however, it is evident that uncertainty is reduced when the uncertainty in AER price is eliminated which can otherwise vary by an order of magnitude. Range is reduced by up to half when variability in AER price is eliminated, and the value is set to $54 per kilogram. This exercise shows that by better defining the production cost of the resin in future work that much of the uncertainty and capture cost can be reduced. After eliminating the variability in these parameters, the range in capture cost is $750–1,300 in the practical scenario.

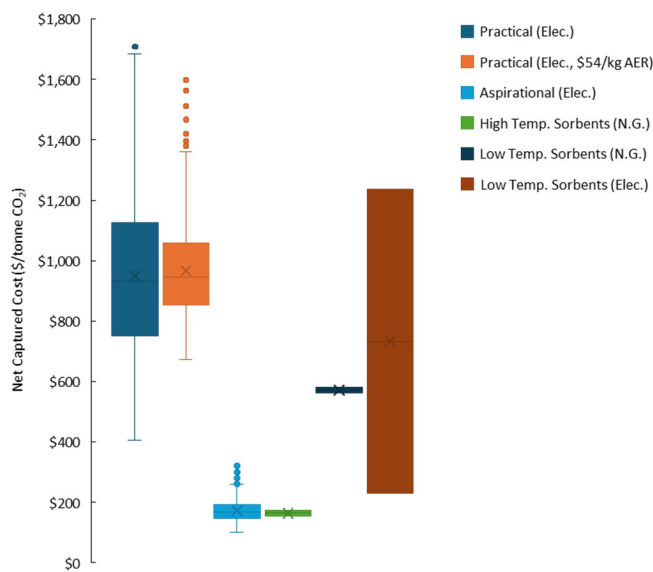

**Figure 10**. Results of Monte Carlo simulation varying inputs which weren't experimentally defined within an expected range.

In addition to energy use, the LCA considers water use. The current process uses ~100 g water per g $CO_2$ captured, which is ~80 fold higher than that 3:1 theoretical molar ratio limit (1.2:1 mass ratio).[20,21] Potential pathways to reduce water uptake are to reduce the ion content in the polymer resin and/or increase the hydrophobicity. The aspirational scenario sets a target for sorbent water uptake of 50 wt.%, which is ~6.6 fold lower than the commercial A501 sorbent. We have synthesized AERs with water uptakes below 50 wt.%, however, they also had lower ion exchange capacities.[50] Dong et. al showed doping AERs with polyvinylidene difluoride (PVDF) could significantly reduce water sorption while maintaining ~0.9 mmol $CO_2$/g capacities.[51]

The results of the sensitivity analysis show us where to focus our attention on improving the system by showing where small changes can have a large impact on results. The results of global sensitivity analysis conducted based on the methods of Morris[42] are presented in **Figure 11** by plotting the mean of the elementary effect and the direction of change across 10 designs in a tornado plot. The mean of the elementary effect is a measure of the influence of each input on cost. The larger the elementary effect the more sensitive the cost is to changes in the input as is the case for the $CO_2$ adsorption rate. Small changes in performance of the AER can lead to large changes in cost. As costs are sensitive to the parameter, future work should focus on increasing it.

In Area 100 ($CO_2$ capture into a brine; Figure 4), adsorption rate, resin lifetime, and resin cost are the most sensitive parameters because they impact conveyor energy use and AER cost. Adsorption rate intuitively impacts the amount of AER required while lifetime and cost impact the replacement costs. A decrease in rate also increases belt length increasing friction between the belt and its track and the energy use of the conveyor. The model is more sensitive to adsorption than desorption because adsorption is an order of magnitude smaller and therefore is the dominant factor in determining delivery rate.

In Area 200 ($CO_2$ extraction from the brine and compression; Figure 4), heat exchanger effectiveness is the most sensitive parameter because it controls heater energy requirements. The direction of change varies for heat exchanger effectiveness depending on the design because there is a local minimum in capture cost in the range of feasible effectiveness values, and therefore cannot be plotted. It is a design variable that controls the tradeoff between heater power and heat exchange size and because the model was relatively sensitive to it, it was crudely optimized and set at 95%. A larger heat exchanger area increases equipment costs but reduces electricity requirements. In general, the collection of variables that determine the energy consumption of Area 200 have a large influence on the model. Other parameters that impact heating energy are the release efficiency, heater efficiency, target bicarbonate molarity, and thermal release temperature. Thermal release efficiency and molarity influence solution volume flow which impact the amount of solution requiring heating. A more efficient $CO_2$ release or a higher molarity solution requires heating smaller volumes that save energy. Intuitively, more energy is required to reach a higher release temperature.

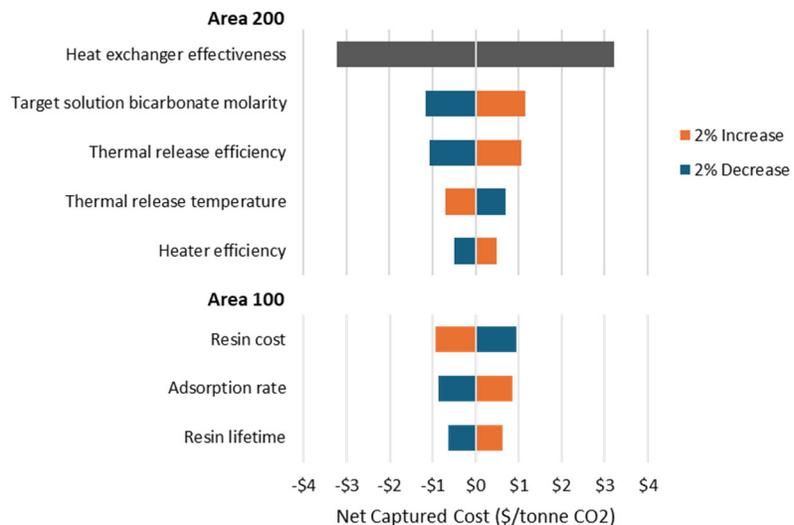

**Figure 11.** Tornado plot of the most sensitive parameters showing the average change in cost across 10 random designs when one input is varies by 2%. Area 100 cover parameters for capturing $CO_2$ into a brine demonstrated in this work and Area 200 covers parameters for extracting $CO_2$ from the brine and compression to 15 MPa modeled from the literature. See also diagram in Figure 4.

**Conclusion**

TEA and LCA reported here show that moisture-driven $CO_2$ capture from passive ambient air flows and delivery into alkaline solutions has significant potential to reduce energy requirements for DAC by up to 87% over processes and sorbents that use heat or vacuum energy to release captured $CO_2$ and fans to move $CO_2$ deplete ambient air streams and the potential to reach $51/tonne for capturing $CO_2$ into alkaline aqueous solutions. Tube mesh packets were developed for containing sorbent powders and beads that maintain a high surface area of the sorbent exposed to air to reduce drying and $CO_2$ loading times 3–4 fold over mesh bags. We envision these packets could be mass-manufactured and loaded at low cost in a roll-to-roll fashion, much like sugar packets and tea bags. Sorbent packets could also be hung from the belt so they could dip into the aqueous solution to keep all moving parts out of the aqueous medium to increase the systems mechanical robustness and longevity. TEA and LCA also demonstrate that further research is also needed to develop sorbents that take up less water to reduce water losses and drying times, thus increasing the amount of $CO_2$ that can be delivered per day, while at the same time increasing $CO_2$ binding capacity, which may require the use of hydrophobic coatings since the $CO_2$ binding sites are hydrophilic. A model was developed that accurately estimates $CO_2$ loading as a function of wind speed under warm and dry conditions that can be used to develop adaptive algorithms to improve $CO_2$ capture and delivery rates. Further quantification of drying and $CO_2$ loading as a function of temperature and RH could improve the model to estimate performance under a wider range of weather conditions. Taken together, with further development, moisture-driven DAC technology can play an important role in capturing and removing $CO_2$ from the atmosphere for use in carbon-based alternatives to fossil-based products or sequestered into durable products or in underground storage to mitigate the impacts from rising levels of atmospheric $CO_2$.

**3. Acknowledgements**

This material is based upon work primarily supported by the U.S. Department of Energy's Office of Energy Efficiency and Renewable Energy (EERE) under the Bioenergy Technology Office award number DE-EE0009274 including design, data collection and analysis. Data analysis and manuscript preparation was also supported by the U.S. Department of Energy, Office of Science, Office of Basic Energy Sciences under Award Number DE-SC0023343. We would also like to thank Kenneth Hodson and Ryan Clark for their help building the impulse sealer to manufacture the sorbent packets.

**4. Author Contributions**

- **Justin Flory**: conceptualization, data curation, formal analysis, funding acquisition, methodology, project administration, supervision, visualization, writing – original draft, writing – review & editing
- **Samantha Taylor**: investigation (system design, build, operation), data curation, formal analysis, methodology, supervision, visualization, writing – review & editing
- **Shuqin Li**: investigation ($CO_2$ delivery), data curation, formal analysis, methodology, visualization, writing – review & editing
- **Sunil Tiwari**: investigation ($CO_2$ delivery), data curation, formal analysis, methodology, visualization, writing – review & editing
- **Garrett Cole**: investigation (technoeconomic and life cycle analyses), data curation, formal analysis, methodology, visualization, writing – review & editing

- **Amory Lowe**: data curation, investigation (control system, $CO_2$ loading model), software, visualization, writing – review & editing
- **Lindsey Hamblin**: investigation (wind tunnel), data curation
- **Samuel Piorkowski**: investigation ($CO_2$ delivery), data curation
- **Matthew Ryan**: conceptualization, investigation (system design, build, operation), writing – review & editing
- **Thiago Stangherlin Barbosa**: investigation (wind tunnel), data curation, formal analysis,
- **Jason Kmon**: investigation (system design, build, operation), writing – review & editing
- **Nick Lowery**: investigation (system design, build, operation), writing – review & editing
- **Joel Eliston**: investigation (system design, build, operation), writing – review & editing
- **Jason Quinn**: conceptualization, formal analysis, funding acquisition, methodology, supervision, writing – review & editing
- **John McGowen**: conceptualization, funding acquisition, methodology, supervision, writing – review & editing
- **Matthew Green**: conceptualization, formal analysis, funding acquisition, methodology, supervision, writing – review & editing
- **Klaus Lackner**: conceptualization, formal analysis, funding acquisition, methodology, supervision, writing – review & editing
- **Wim Vermaas**: conceptualization, formal analysis, funding acquisition, methodology, project administration, supervision, writing – review & editing

## 5. Conflicts of Interest

The authors declare no conflicts of interest related to the content of this paper. M.D.G. is co-founder of NuAria, LLC and owns 50% equity interest in the company. The work of this manuscript is not directly related to the activities of the company. K.S.L is an advisor to CarbonCollect Limited and DACLab, which are both DAC companies, and to Aircela Inc., which aims to produce fuel from DAC $CO_2$ and renewable energy. Arizona State University has licensed part of its DAC intellectual property to Carbon Collect Limited and owns a stake in the company. As an employee of the University, K.S.L. is a technical advisor to the company and in recognition also received shares from the company. Carbon Collect Limited also supports DAC research at Arizona State University. K.S.L. also has a financial stake in Aircela and is on the Board of Aircela. None of these companies are involved in the work reported here.

## 7. Supplementary Information

***100g* system with one continuous loop.** The initial *100g* design had a single loop with V shaped path to increase the fraction of belt exposed to the air for drying and $CO_2$ loading (**Figure S1**). A schematic showing the different belt path of the single continous loop vs two independent loops is shown in **Figure S2**.

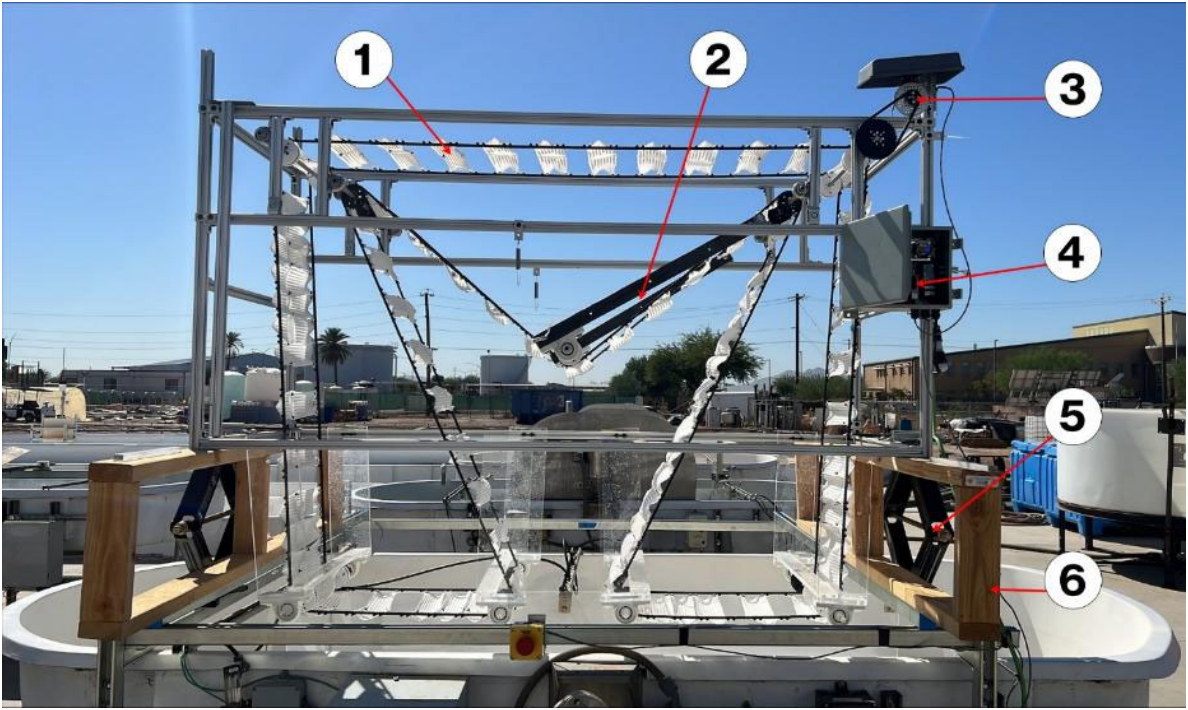

**Figure S1:** *100g* unit installed at AzCATI seen in the lifted "Up" position. 1) Double-wide packet belt; 2) Tensioning axle; 3) Servo motor with rain shade; 4) Electronics in a weather box; 5) Car jacks to lift the system; and 6) Safety braces.

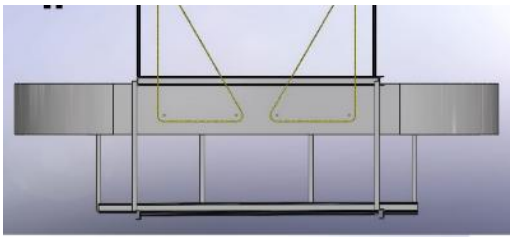
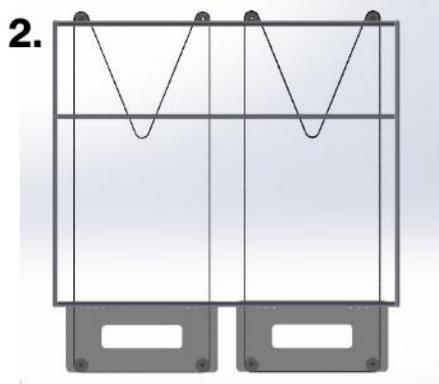

**Figure S2.** Schematic showing the different belt path of the single continous loop (top) vs two independent loops (bottom) used in the first (top) and second (bottom) generation *100g* systems.

The mesh packets were made using a custom-built impulse sealer (**Figure S3A**) comprised of a foot pedal actuator linked to a pneumatic line, working as a heat press with individual impulse seal elements that heat in sequence. This creates an eight-stick tube packet that only requires a few additional seals by hand to prepare for sorbent loading. An image of a double wide sorbent packet connected to parallel chains is shown in **Figure S3B**.

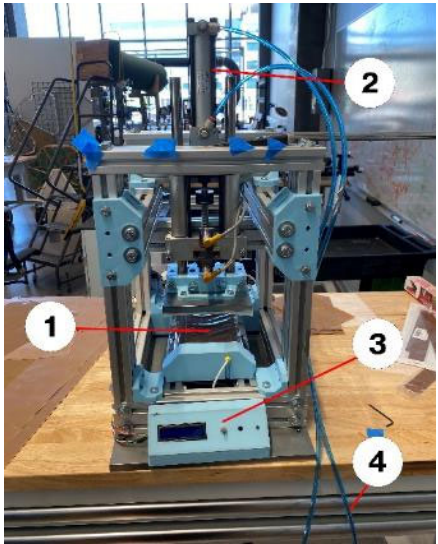

**Figure S3A:** Impulse heat press. 1) Nine parallel impulse elements, 2) pneumatic linear actuator, 3) dwell time control, and 4) pneumatic line to foot pedal.

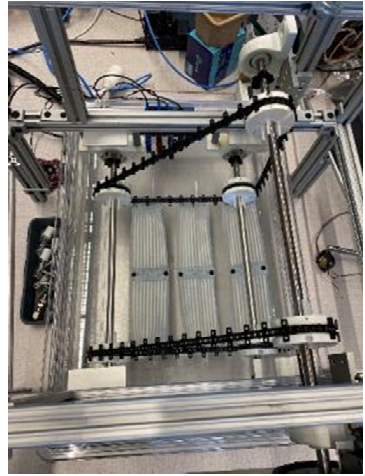

**Figure S3B:** 1g system showing how two packets are snapped together and fastened to parallel chains for rotating around the track.

**Water uptake of A501, Excellion and nylon mesh.** The water uptake and drying of a 1.6 g Excellion sheet, and nylon mesh packet (1.3 g) with 0.6 g of sorbent as well as an empty mesh packet (2.0 g) as measured on a Futek load cell (Figure S4A) is shown as a fraction of the dry weight as averages of 4 replicated wet/dry cycles in **Figure S4B**. The steeper decline during the first 5 min for the mesh packets with A501 or empty is due to excess water dripping from the mesh material after coming out of the water bath. In a normal system this water would drip back into the water source so excluded from the reported water uptake values of 330 wt.% for A501 mesh packet and 57 wt.% for the empty mesh packet.

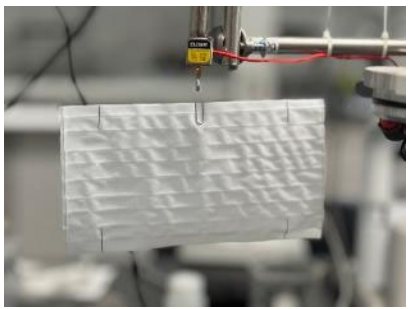

**Figure S4A.** Futek load cell used to measure the weight change over time of an A501 loaded sorbent packet.

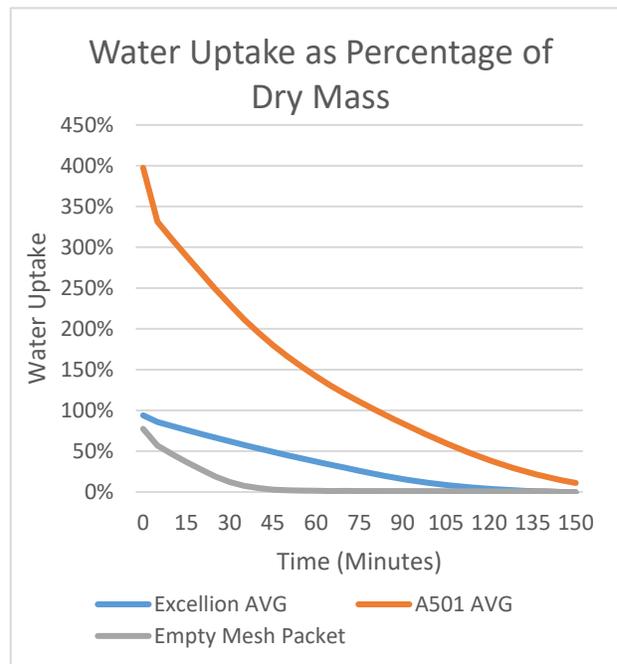

**Figure S4B.** Water uptake and drying in ambient still lab air of Excellion sheets, A501 loaded mesh packet and empty mesh packet.

**Technoeconomic and life cycle analysis (TEA/LCA) calculations and assumptions.**

This section documents the engineering process model. Key variables in the model are based on experimental data when available. Experimental values of the key variables are provided in **Table S1** for both the aspirational and practical scenarios.

**Table S1: Values varied between the aspirational and practical scenarios**

| Variable | Units | Value | | Reference |
|---|---|---|---|---|
| | | Practical | Aspirational | |
| Adsorption rate ($R_{sorp}$) | µmol/g-s | 0.08 | 0.125 | Experiment |
| Desorption rate ($R_{desorp}$) | µmol/g-s | 0.5 | 2 | Experiment |
| Effective $CO_2$ binding capacity ($ix$) | µmol/g | 675 | 4,000 | X. Wang[46] |
| Water uptake of resin ($f_w$) | g/g resin | 3 | 0.5 | Experiment |
| Resin lifetime | year | 5 | 10 | Assumption |

Other variables were held constant between the two scenarios and were defined with a large range in value to account for uncertainty due to a lack of experimental research. Their values are provided in **Table S2** and **Table S3**.

**Area 100: Resin cycling**

The following sections outline the main calculations of the model of the anion exchange resin, resin for short, conveyance system. Inputs for Area 100 that weren't experimentally calculated are provided in **Table S2**.

**Table S2: Resin cycle variables**

| Variable | Units | Value | | Reference |
|---|---|---|---|---|
| | | Min. | Max. | |
| Mass ratio of resin to resin support material ($f_s$) | | 1 | 10 | Assumption |
| Conveyor load rating ($L_T$) | kg/m | 178 | 267 | |
| Resin sheet length | m | 0.1 | 0.3 | Assumption |
| Resin sheet thickness ($t$) | m | 0.03 | 0.15 | Assumption |
| Resin bulk density ($rho_{rsn}$) | kg/m3 | 560 | 960 | Wheaton[52] |
| Max conveyor height ($h$) | m | 3 | 8 | Assumption |
| Friction factor between conveyor and track ($\mu$) | | 0.025 | 0.035 | ISO 9001[53] |
| Mass of conveyor ($m_B$) | kg/m | 1 | 10 | L.K. Goodwin Co.[54] |
| Mass ratio of conveyor support to load | kg support/kg load | 0.03 | 0.3 | Assumption |

**Anion exchange resin**

The system was sized based on the delivery rate. The theoretical delivery rate is the fastest rate at which $CO_2$ can be delivered to the solution. To calculate theoretical delivery rate ($R_{delivery}$) in micromoles per gram per second, it is assumed that no time is wasted in transition between stages of the cycle. In other words, the instant the polymer adsorbs a specified amount of $CO_2$, it enters the solution, the instant it desorbs a specified amount of $CO_2$, it begins to dry, and the instant it dries, it begins to adsorb $CO_2$ again completing the cycle. It is calculated by Equation S1 where $R_{sorp}$ is the average adsorption rate and $R_{desorp}$ is the average desorption rate to adsorb and desorb the specified amount of $CO_2$ in micromoles per gram per second. The specified amount of $CO_2$ absorbed and desorbed does not need to be equal the $CO_2$ binding capacity. Thresholds can be set so that desorption and adsorption stop before all the ion exchange sites are exchanged which can avoid the slower adsorption and desorption rates experienced after most sites have already been exchanged. In future work, these thresholds can be optimized to maximize delivery rate. In this work delivery rate was calculated, assuming the threshold was 90% of the $CO_2$ binding capacity.

$$R_{delivery} = \frac{1}{\frac{1}{R_{sorp}} + \frac{1}{R_{desorp}}}$$ **Equation S1**

In practice, the rate at which $CO_2$ is delivered to the solution is lower than the theoretical delivery rate because, given a system without dynamic feedback, the cycle timing cannot be adjusted on demand to move the resin from the solution to the air or vice versa the instant it is required. Wind speed, solution pH, temperature, and humidity will all impact adsorption and desorption rates which could lead to imperfect cycle timing decreasing the actual delivery rate. The delivery rate determines how big a system and how much resin is required. The amount of resin required by the system ($Rsn$) in kilograms is calculated by Equation S2 where $D_{CO_2}$ is the required delivery rate in micrograms per second, $R_{delivery}$ is the delivery rate of the resin in micromoles per second per gram of resin, and $M_{CO_2}$ is the molar mass of $CO_2$.

$$rsn = \frac{D_{CO_2}}{M_{CO_2} R_{delivery}} 10^{-3}$$ **Equation S2**

**Power usage of the conveyor**

While delivery rate determines the size of the system, binding capacity determines the energy usage of the system. The energy for the belt was modeled based on a Z-Type (Zig-Zag) overhead chain conveyor (No. 2035 from L.K. Goodwin Co.).[54] The energy used in driving the belts is approximated from three components, the friction between the conveyor and its track, the drag on the resin in the solution, and the energy for lifting the additional mass of water absorbed by the resin out of the solution. The total length of the belt and the number of cycles it make annually will prove important in calculating energy lost to friction and drag and can be calculated by Equation S3 and Equation S4 where $L_T$ is the capacity of the conveyor in kilograms per meter, $f_w$ is the water uptake as a fraction of resin mass, $f_s$ is the amount of structural plastic as a fraction of resin mass and $ix$ is the $CO_2$ binding capacity micromoles per gram.

$$L_B = \frac{rsn\,(1 + f_w + f_s)}{L_T}$$ **Equation S3**

$$N = \frac{R_{delivery}\,3.1536 * 10^7}{ix}$$ **Equation S4**

To calculate drag, assumptions about the geometry of the resin in water needed to be made. For illustrative purposes, it was assumed the resin would take form of flat sheets. The fraction of resin moving through the water at any given time is key to determining drag force. It is calculated by Equation S5 and is equal to the ratio of adsorption rate to desorption rate. Drag coefficient for a flat plate ($C_D$) is calculated in Equation S6 and is a function of Reynolds number where the characteristic dimension is the length of the sheet in the direction parallel to the flow.[53] Drag force is calculated using Equation S7 where $rho_w$ is the density of water in kg/m³, $V$ is the belt speed in m/s which can be optimized though the number of conveyor circuits implemented, and $A$ is the total surface area of all the sheets in square meters which is a function of the chosen dimensions and bulk density of the flow.[53] Area can be calculated in Equation S8 where $t$ is the resin thickness and $rho_{rsn}$ is the bulk density of the resin. Energy in kilojoules per year is ultimately calculated from the force and the total distance on which the force acts over the course of a year using Equation S9 where $L_B$ is the total length the belts move per cycle.

$$f = \frac{R_{sorp}}{R_{desorp}}$$ **Equation S5**

$$C_D = \frac{1.328}{Re^{\frac{1}{2}}}$$ **Equation S6**

$$F_d = \frac{C_D}{2} rho_w V^2 (A\,f)$$ **Equation S7**

$$A = \frac{2\,rsn}{t\,rho_{rsn}}$$ **Equation S8**

$$E_d = N\,F_d\,(L_B\,f)\,10^{-3}$$ **Equation S9**

Energy lost to friction between the belt and the track in kilojoules per year was calculated from the sum of the masses of resin, belt, water and resin structural backing, and a coefficient of friction in Equation S10 where $\mu$ is the coefficient of friction (0.025 base on manufacturer guidance), $g$ is the gravitational constant (9.81 m/s²), and $m_B$ is the mass of the conveyor in kg per meter.

$$E_f = \mu\,(rsn\,(1 + f_w + f_s) + L_B\,m_B)\,g\,L_B\,N\,10^{-3}$$ **Equation S10**

The energy for lifting in the additional mass of water absorbed by the resin out of the solution in kilojoules per year was calculated using Equation S11 where a height ($h$) in meters was assumed for the height of the overhead conveyor in meters.

$$E_g = rsn\, f_w\, N\, g\, h\, 10^{-3} \quad \textbf{Equation S11}$$

**Equipment**

The remaining variables help to define the amount and size for conveyor belt components and their prices. The size of the motors is based on power requirements. The length of conveyor is based on how much resin can be carried per length of belt which is a function of water uptake by the resin and the mass of non-resin material used to affix the resin to the belt. The cost of belt and track are based on belt length. The cost of supports was estimated based on the weight of the track, belt, and mass conveyed by the belt. The amount of non-resin material affixing the resin to the belt was estimated based on the mass of resin using the variable "Mass ratio of resin support material to resin" ($f_s$).

**Area 200: Thermal release column**

The following sections outline the main calculations of the model of the thermal release column. Inputs for Area 200 that weren't experimentally calculated are provided in **Table S3**.

**Table S3: Thermal release column variables**

| Variable | Units | Value Min. | Value Max. | Reference |
|---|---|---|---|---|
| Target solution bicarbonate molarity ($M$) | M | 1 | 1.2 | Assumption |
| Thermal release efficiency ($\xi$) | g release/g theoretical | 0.5 | 1 | Assumption |
| Storage time | days | 1 | 3 | Assumption |
| Heat exchanger effectiveness ($\varepsilon$) | | 0.95 | 0.95 | Assumption |
| Thermal release temperature ($T_{op}$) | C | 95 | 95 | Assumption |
| Ambient temperature ($T_{amb}$) | C | 20 | 30 | Assumption |
| Heat transfer coefficient ($U$) | W/m2-K | 350 | 450 | Assumption |
| Heater efficiency ($\eta$) | | 1 | 1 | Assumption |
| Thermal release residence time | s | 200 | 400 | Assumption |
| Compressor size | MWh/tonne $CO_2$ | 0.12 | 0.13 | Assumption |

**Solution circulation volume**

The resin cycling process delivers $CO_2$ to an alkaline media solution. The $CO_2$ is recovered as pure gas by the thermal release column. The mass of solution circulating in the system in kilograms per second is defined by Equation S12 where M is the molarity of the solution in mols per liter, 2 is the moles of bicarbonate produced per mole of $CO_2$, $\rho$ is the density of the solution in kilograms per cubic meter, and $\xi$ is the fraction of $CO_2$ stored in solution that is liberated in the release column.

$$m_s = \frac{2 \cdot 10^{-9}\, D_{CO_2}\, \rho_s}{M_{CO_2}\, M\, \xi} \quad \textbf{Equation S12}$$

The performance of the Thermal Release Column is still unverified, but preliminary estimates suggest roughly half of the carbon stored in the bicarbonate solution is extracted by heating the fluid ($\xi = 0.5$). The size of the reactor is based on an expected solution residence time and required flow rate.

**Energy for heating**

The energy for operating the column in megajoules per year was calculated by Equation S13. The solution is heated from 20 °C to the column operating temperature before entering the column. A heat exchanger is used to preheat the solution to by recovering heat from the solution exiting the column. An electric emersion heater is assumed for heating in order to reduce greenhouse gas emissions.

$$E_{col} = \eta\, m_s\, C\, (T_{op} - T_{in})\, 31.536 \quad \textbf{Equation S13}$$

Where $m_s$ is the mass of circulating solution in kilograms per second defined by Equation S12, $\eta$ is the efficiency of the heater (100% for an electric emission heater), $C$ is the specific heat capacity of the solution in joule per kilogram kelvin, $T_{op}$ is the column operating temperature (heater outlet temperature) in °C, and $T_{in}$ is the heater inlet temperature (heat exchange cold side outlet temperature) in °C.

Heater inlet temperature is related to the exchanger design and is calculated in Equation S14 where $\varepsilon$ is the heat exchanger effectiveness and $T_{amb}$ is the ambient temperature in °C.

$$T_{in} = \varepsilon\, (T_{op} - T_{amb}) + T_{amb} \quad \textbf{Equation S14}$$

There is a tradeoff between heat exchange size and corresponding capital investment and heater energy usage and corresponding operating costs. Heat exchange effectiveness is a design variable that controls this tradeoff and because the model was relatively sensitive to it, it was crudely optimized and set at 95%. Heat exchange area was calculated by Equation S15 where NTU is defined in Equation S16 and U is the overall heat transfer coefficient.

$$A = \frac{NTU\, m_s\, C}{U} \quad \textbf{Equation S15}$$

$$NTU = \frac{\varepsilon}{1-\varepsilon} \quad \textbf{Equation S16}$$

**Energy use of the compressor**

The power of the compressor was derived from Keith et al. as the power to compress and dry $CO_2$.[47]
Equipment The remaining variables help to define the equipment costs. Time in storage and residence time in the column help determine the size of the tanks and columns.

**Life cycle assessment**

Carbon intensities (CI) are provided in **Table S4**.

**Table S4: Cabon intensities (CI).**

| Material | CI | Reference |
|---|---|---|
| Electricity | 0.1–0.695 kg $CO_2$e/kWh | U.S. Environmental Protection Agency eGRID |
| Resin | 2.56–3.84 kg $CO_2$e/kg | USLCI, Amini[55] |
| Process water | Ecoinvent database | |
| Sodium bicarbonate | Ecoinvent database | |
| Steel | Ecoinvent database | |
| Structural plastic | 1–5 kg $CO_2$e/kg | Assumption |

**Techno-economic assessment**

Manufacturing costs are derived from the amounts in the life cycle inventory and the material prices provided in **Table S5** with the exception of labor costs which were modeled on the same assumptions and labor requirements as Keith et al. ($30 per tonne of $CO_2$ captured assuming 90% utilization).[47] Conveyor costs were assumed to scale linearly and their reference cost are provided as a unit cost (e.g. $/kWh). Equipment used in the model is listed in **Table S6**. Costs for each component were calculated that are representative of the size equipment used in most simulation runs and are scaled by a scaling exponent of 0.8 when a simulation run requires a slightly smaller or larger piece of equipment.

**Table S5: Material and energy price data**

| | | Price | | Reference |
|---|---|---|---|---|
| | | min | max | |
| Resin | $/kg | 7 | 100 | Sinhai[55] |
| Structural plastic | $/kg | 1 | 10 | Assumption |
| Process water | $/L | 0.00022 | 0.00036 | Assumption |
| Sodium bicarbonate | $/kg | 0.3 | 0.6 | Assumption |
| Electricity | $/kWh | 0.04 | 0.16 | U.S. Energy Information Administration |

**Table S6: Equipment cost data**

| Equipment | Capacity/ Amount | Price min | Price max | Reference |
|---|---|---|---|---|
| Area 100 | | | | |
| Storage tank | # m$^3$ | $430,000 | $640,000 | |
| Track | 1 m | $56 | $84 | L.K. Goodwin Co.[54] |
| Chain | 1 m | $130 | $190 | L.K. Goodwin Co.[54] |
| Drive | 1 kW | $9,200 | $14,000 | L.K. Goodwin Co.[54] |
| Support | 1 kg | $2 | $20 | Assumption |
| Area 200 | | | | |
| Compressor | 2,000 kW | $490,000 | $730,000 | Towler[56] |
| Thermal release tank | 43 m$^3$ | $22,000 | $34,000 | Towler[56] |
| Heater cost | 150 kW | $4,800 | $7,200 | Towler[56] |
| Heat exchanger cost | 1,000 m$^2$ | $190,000 | $290,000 | Towler[56] |

## Results
### Life cycle inventory

The life cycle inventory for a tonne of $CO_2$ captured is provided in **Table S7**. Notably, these results do not reflect the amount of material to capture a net tonne.

**Table S7: Life cycle inventory**

| Material | Units | Practical | Aspirational |
|---|---|---|---|
| Operation | | | |
|   Area 100 | | | |
|     Electricity | kWh | 160–800 | 6.2–38 |
|     AER | Kg | 2.1 | 0.79 |
|     Structural plastic | Kg | 0.21–2.1 | 0.079–0.79 |
|     Water | L | 100,000 | 5,700 |
|   Area 200 | | | |
|     Electricity | kWh | 190–530 | 190–530 |
|     Water | L | 5.2–1,500 | 5.2–1,500 |
|     Sodium bicarbonate | Kg | 0.36–63 | 0.36–63 |
| Construction | | | |
|   Area 100 | | | |
|     AER | Kg | 0.39 | 0.29 |
|     Structural plastic | Kg | 0.15–1.4 | 0.035–0.33 |
|     Steel | kg | 0.039–0.39 | 0.029–0.29 |

### Life cycle impact analysis

In the practical scenario, 0.14–0.60 tonnes of $CO_2$-e are emitted per tonne of $CO_2$ captured with 95% confidence. There is, however, and unlikely possibility that the practical scenario emits more $CO_2$-e than it captures if the upper end of the range in the LCI is assumed for every material or energy amount. In the aspirational scenario, this is reduced to 0.06–0.29.

**Capital investment**
The capital investment is shown in **Table S8** and is based on a facility with a capacity of 100,000 tonnes of $CO_2$ Capture per year. The Amount of $CO_2$ net captured is less and is dependent of process emissions. Site development, Additional piping, and warehouse cost were assessed as a percentage of installed equipment cost. Indirect costs were assessed as a percentage of total direct costs. Land was assumed at 1 Acre for $100,000. Working capital was assess as a percentage of the fixed capital investment

**Table S8: Capital investment**

| Item | Practical | Aspirational |
|---|---|---|
| Direct costs | | |
|   Area 100 | $124,029,000 | $6,255,000 |
|   Area 200 | $25,239,000 | $10,248,000 |
|   Warehouse | $5,970,723 | $660,000 |
|   Site development | $13,434,127 | $1,485,000 |
|   Additional Piping | $6,717,063 | $742,000 |
| Indirect costs | | |
|   Proratable expenses | $17,539,000 | $1,939,000 |
|   Field expenses | $17,539,000 | $1,939,000 |
|   Project contingency | $35,078,000 | $3,878,000 |
|   Home office and construction fee | $17,539,000 | $1,939,000 |
|   Other costs (start-up, permits, etc.) | $17,539,000 | $1,939,000 |
| Working Capital | $14,031,000 | $1,551,360 |
| Land | $100,000 | $100,000 |

**Operational Expenditures**

**Table S9: Operation expenditures**

| Item | Practical | Aspirational |
|---|---|---|
| Fixed costs | | |
|   General | | |
|     Labor | $2,700,000 | $2,700,000 |
|   Area 100 | | |
|     Material replacement | $11,879,000 | $78,000 |
|   Area 200 | | |
|     Sodium bicarbonate | $545,000 | $544,000 |
|     Process Water | $9,000 | $8,000 |
| Variable costs | | |
|   Area 100 | | |
|     Electricity | $2,470,000 | $6,600 |
|     Process water | $2,636,000 | $74,000 |
|   Area 200 | | |
|     Electricity | $4,119,000 | $2,637,000 |

**Measured CO$_2$ loading vs sorbent form factor**. The A501 sorbent CO$_2$ loading as a function of time (Figure 5) data were used to determine the maximum CO$_2$ uptake (**Figure S5A**) and the maximum CO$_2$ capacity (**Figure S5B**) when the sorbent was contained within elongate mesh tube packets compared to a mesh bag with a single compartment, which led to uneven setting (**Figure S6**).

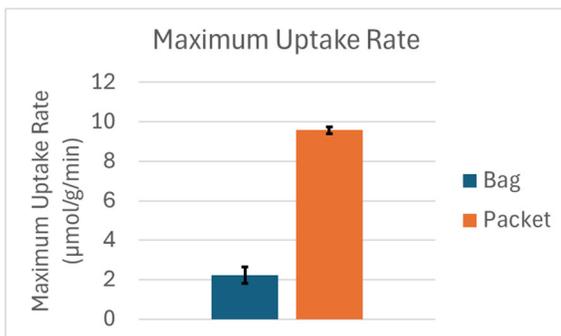

**Figure S5A.** Maximum CO$_2$ uptake rate of A501 contained in a mesh packet vs mesh bag.

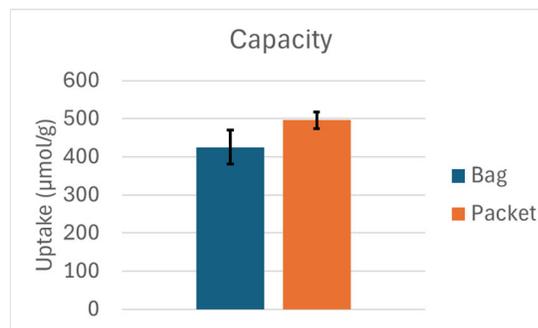

**Figure S5B.** A501 CO$_2$ capacity when contained in a mesh packet vs mesh bag.

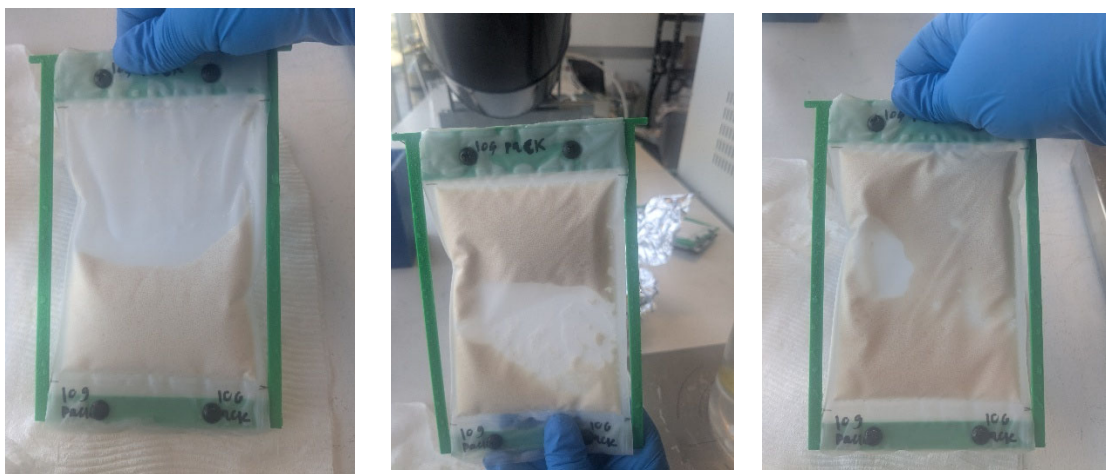

**Figure S6.** Images of the A501 sorbent beads contained within a single mesh bag after being immersed in water (left), and after two different attempts to disperse the beads (center and right).

**1g system with Excellion AER sheets.** The first *1g* system was evaluated with 40 Excellion AER sheets (~8" long x 11/16" wide) attached to a chain driven belt as shown in **Figure S7** and the basin was filled with ~12 L of 1 mM $K_2CO_3$. Initially the system was configured to run a complete wet / dry cycle every 1.5 h. As shown in **Figure S8A** the pH of the system with sorbent installed drops much more rapidly than a control without sorbent until the solution becomes saturated and sorbent delivered $CO_2$ is in equilibrium with $CO_2$ outgassing to ambient air at pH ~8.2. NaOH was added after 20 and 48 hrs to reset the pH and increase the buffer capacity to store additional $CO_2$ which enabled the system to deliver similar amounts of $CO_2$ each time.

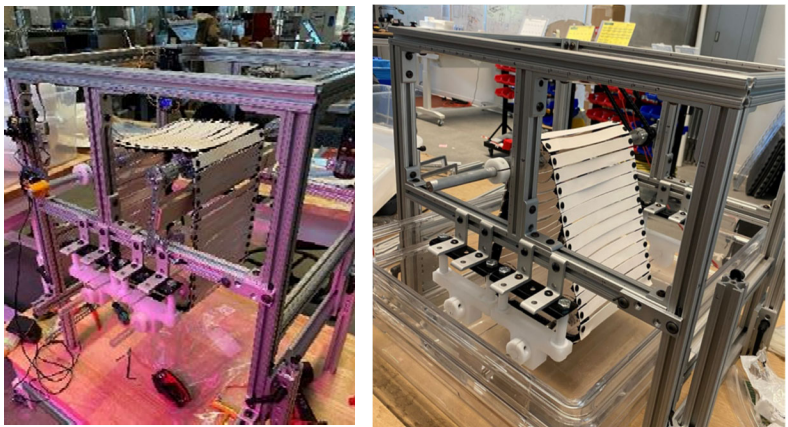

**Figure S7.** Photo of the *1g* system with flat sheets of Excellion ion exchange resin arranged with four axels in a rectangle (left) or three axels in a triangle (right).

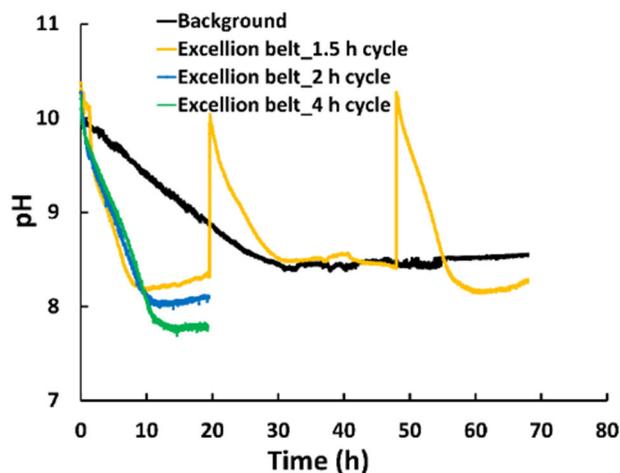

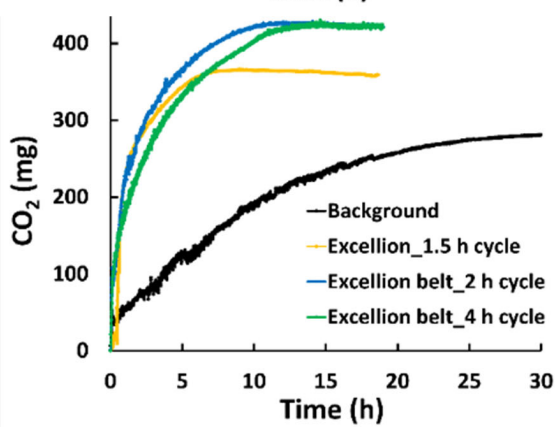

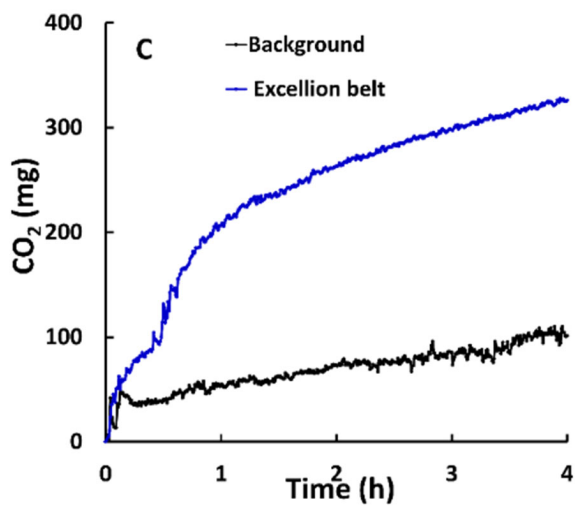

**Figure S8.** Comparison of pH kinetics (reflecting $CO_2$ delivery) (A) and released $CO_2$ (B) into 12 L of 1 mM $K_2CO_3$ solution by an Excellion belt with 1.5 h, 2 h and 4 h cycle time. (C) shows the average $CO_2$ delivered within the first 4 h of operation with 1.5 h, 2 h and 4 h cycle time.

**CO₂ delivery by the *100g* system into a 4.2 m² raceway pond. Figure S9** shows inorganic carbon data from a pond (#28) connected to the *100g* system compared to a control pond (#20). Initially when the system was turned on (April 15) it had 188 two-gram packets; 50 more were installed on April 19 and another 40 on April 26. The system was stopped periodically due to mechanical issues before having a nice continuous run from April 30 to May 6, which showed anticipated levels of $CO_2$ delivery (**Figure 8**).

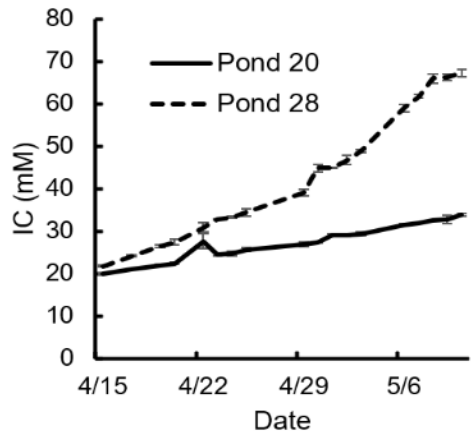

**Figure S9.** IC analysis of the medium from the *100g* system pond (#28) and a control pond (#20).

**Measured and estimated CO₂ loading vs wind speed.** The combined sorbent $CO_2$ loading rates were measured in the wind tunnel at air flowrates of ~0.7 m/s (lower limit of the wind tunnel), 1.6 m/s (average wind speed in Phoenix) and 5 m/s (**Figure S10**). These data were fit with a logistic three parameter (3P) model (**Equation S17**) to determine a $CO_2$ loading level as a function of time at each wind speed:

$$CO_2 \text{ loading } (t) = \frac{c}{1+e^{-a(t-b)}} \quad \textbf{Equation S17}$$

where $c$ is the maximum observed $CO_2$ loading in μmol $CO_2$ per g of A501 sorbent, $a$ is the exponential decay term in units of per hour for a specific wind speed, $t$ is the amount of time in hours the sorbent has been in contact with the wind and $b$ is the time when the sorbent has loaded to

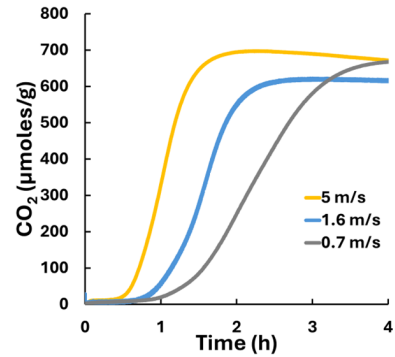

**Figure S10.** A501 sorbent drying and $CO_2$ loading from air contained within a mesh tube packet as a function of wind speed (15% relative humidity, 22 °C).

50% of its maximum $CO_2$ capacity observed as the inflection point in the curve where the exponential decay in the loading rate begins (see **Figure S6**). The exponential decay of the $CO_2$ loading rate fitting parameter ($a$) determined from Equation S17 was plotted for each measured wind speed ($W$) and fitted to a power model (**Figure S7**) described in **Equation S18** and used to estimate $CO_2$ loading level as a function of the time that a packet was exposed to the wind at a measured average wind speed,

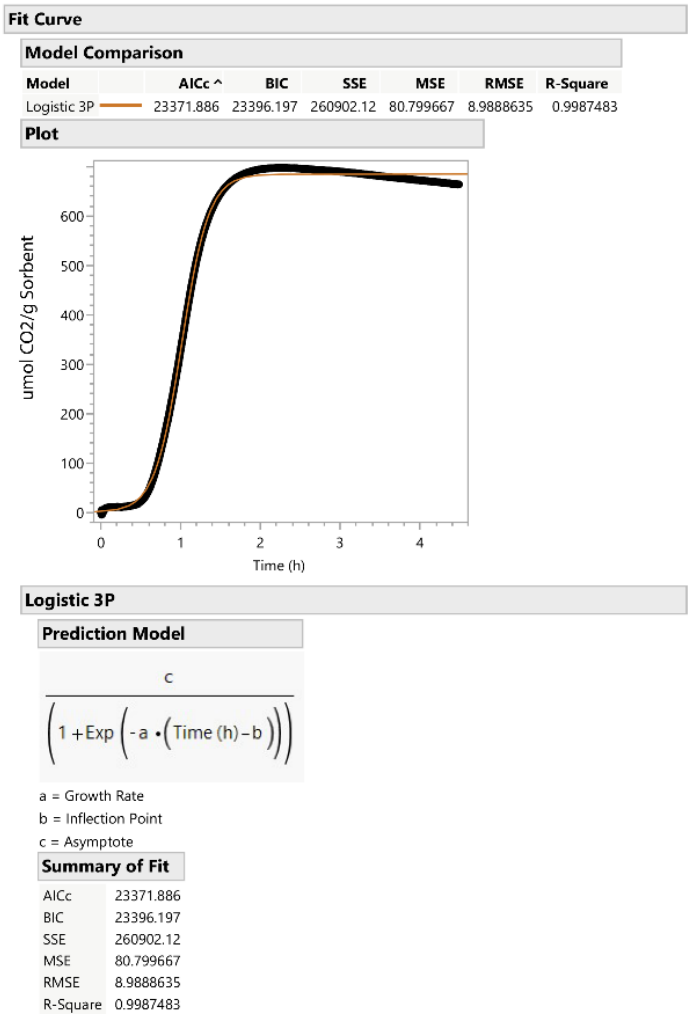

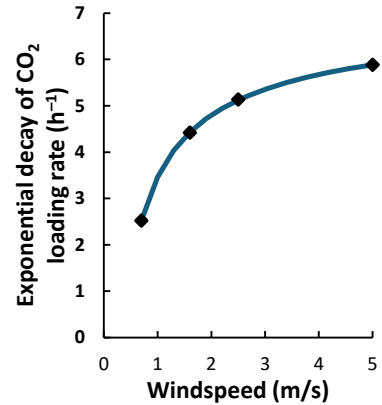

**Figure S12.** Power model fit (1.6% RMS error) and projection to higher wind speeds of the average $CO_2$ loading rate of the mesh tube packets as a function of wind speed.

**Figure S11.** Statistical curve fit of initial wind tunnel data showing mathematical model of $CO_2$ uptake for 5 m/s windspeed. The estimation model for this serves as the mathematical model used to calculate estimated $CO_2$ uptake in individual belt sections.

Exponential decay of $CO_2$ loading rate $(W) = d + f \times W^g$     **Equation S18**

where $d$ is the wind speed at zero exponential decay, $f$ is a scaling factor based on the rate at which the exponential decay changes as a function of windspeed ($W$) in m/s and $g$ is the observed power modifier that determines how strongly the exponential decay of the loading rate is dependent on $W$. The root mean squared (RMS) error of the fit in **Figure S7** was 1.6%. The process was also used to model the inflection point ($b$) (data not shown) using **Equation S18**.

A model was developed to estimate the $CO_2$ loading rate of the sorbent packets as a function of wind speed using **Equation S18** to determine the exponential decay of the $CO_2$ loading rate ($a$) and inflection point ($b$) as a function of wind speed ($W$) and plugged into Equation S17 to determine the total $CO_2$ loaded on a sorbent exposed to an average wind speed ($W$) for a specified amount of time in hours ($t$), as shown in **Equation S19**. The numerical values shown are model fitting parameters.

$$CO_2 \text{ loading } (W, t) = \frac{670}{1+e^{-\left(7.337+(-3.874*(W)^{-0.6098})*\left(t-\left(0.588+1.285((W)^{-0.7})\right)\right)\right)}} \quad \textbf{Equation S19}$$